\documentclass[12pt]{article}

\usepackage{booktabs}   
\usepackage{array}      
\usepackage{xcolor}     
\usepackage{tikz}       
\usepackage{tabularx}
\usetikzlibrary{positioning, arrows.meta}
\usepackage[most]{tcolorbox} 
\definecolor{sosblue}{HTML}{1F4E79}     
\definecolor{sosbluemid}{HTML}{2A78D6}  
\definecolor{sosbluepale}{HTML}{EAF1F9} 
\definecolor{sosgray}{HTML}{5A5A5A}     
\definecolor{soscat1}{HTML}{2A78D6} 
\definecolor{soscat2}{HTML}{EB6834} 
\definecolor{soscat3}{HTML}{1BAF7A} 
\definecolor{soscat4}{HTML}{EDA100} 
\definecolor{soscat5}{HTML}{E87BA4} 

\newcommand{\soscol}[1]{\textcolor{sosblue}{\textsf{\small #1}}}

\newtcolorbox{rowcard}[2][]{%
  enhanced, colback=sosbluepale, colframe=sosblue,
  boxrule=0.6pt, arc=1.5mm, left=2mm, right=2mm, top=1mm, bottom=1.5mm,
  fonttitle=\bfseries\small\sffamily, coltitle=white,
  colbacktitle=sosblue, title={#2}, #1}

\newtcolorbox{fieldnote}[1]{%
  colback=white, colframe=sosblue!60, boxrule=0.5pt, arc=1mm,
  left=2mm, right=2mm, top=1mm, bottom=1mm,
  fonttitle=\bfseries\sffamily\small, coltitle=sosblue,
  title={#1}, attach title to upper={\ \textcolor{sosgray}{---}\ }}

\newcommand{\chip}[1]{\tikz[baseline=(c.base)]{%
  \node[draw=sosblue!70, fill=sosbluepale, rounded corners=2pt,
        inner xsep=3pt, inner ysep=1.5pt, font=\scriptsize\sffamily] (c) {#1};}}
\newcommand{\chiphl}[1]{\tikz[baseline=(c.base)]{%
  \node[draw=soscat2, fill=soscat2!15, rounded corners=2pt,
        inner xsep=3pt, inner ysep=1.5pt, font=\scriptsize\sffamily\bfseries] (c) {#1};}}

\newcommand{\soscode}[2]{\tikz[baseline=(c.base)]{\node[fill=#1, text=white, rounded corners=2pt,
  minimum width=7mm, inner ysep=1.5pt, font=\sffamily\bfseries\scriptsize] (c) {#2};}}
\newtcolorbox{sosprompt}[1]{enhanced, colback=white, colframe=sosblue!60, boxrule=0.5pt,
  arc=1mm, left=2mm, right=2mm, top=1.5mm, bottom=1.5mm, fontupper=\small,
  fonttitle=\bfseries\sffamily\small, coltitle=sosblue, colbacktitle=sosbluepale, title={#1}}
\newtcolorbox{soscodebook}[2]{enhanced, colback=sosbluepale, colframe=sosblue, boxrule=0.6pt,
  arc=1.5mm, left=2mm, right=2mm, top=1.5mm, bottom=1mm, fontupper=\small,
  fonttitle=\bfseries\sffamily\small, coltitle=white, colbacktitle=sosblue,
  title={#2\hfill\textmd{\textit{#1}}}}
\newtcolorbox{sosscale}[1]{enhanced, colback=white, colframe=sosblue!60, boxrule=0.5pt,
  arc=1mm, left=2mm, right=2mm, top=1.5mm, bottom=1.5mm, fontupper=\small,
  fonttitle=\bfseries\sffamily\small, coltitle=sosblue, colbacktitle=sosbluepale, title={#1}}
\newtcolorbox{sosrules}[1]{breakable, enhanced jigsaw, colback=white, colframe=sosblue!60,
  boxrule=0.5pt, arc=1mm, left=2mm, right=2mm, top=1.5mm, bottom=1.5mm, fontupper=\small,
  fonttitle=\bfseries\sffamily\small, coltitle=sosblue, colbacktitle=sosbluepale, title={#1}}

\usepackage[utf8]{inputenc}
\usepackage[margin=1in]{geometry}
\usepackage[auth-sc, affil-it]{authblk}
\usepackage{graphicx}
\usepackage{booktabs}
\usepackage[sort&compress, round]{natbib}
\usepackage[doublespacing]{setspace}
\usepackage{longtable}
\usepackage{booktabs}
\usepackage{siunitx}
\newcolumntype{d}{S[
    input-open-uncertainty=,
    input-close-uncertainty=,
    parse-numbers = false, 
    table-align-text-pre=false,
    table-align-text-post=false
 ]}
\usepackage{caption}
\usepackage{lipsum}
\usepackage{scrextend}
\usepackage{natbib}
\usepackage{setspace,lipsum}
\usepackage{chngpage}
\usepackage{lscape}
\usepackage{amssymb}
\usepackage{siunitx}
\usepackage{dcolumn}
\usepackage{subfig}
\usepackage{soul,color}
\usepackage{amsmath}
\usepackage{verbatim}
\usepackage{epigraph}
\usepackage{breakcites}
\usepackage{dirtytalk}
\usepackage{graphicx}
\usepackage[affil-it]{authblk}
\graphicspath{ {./images/} }
\usepackage{float}
\usepackage{appendix}
\usepackage{caption}
\usepackage{titlesec} 
\usepackage{verse}
\usepackage{array}
\usepackage{csquotes}
\usepackage{wrapfig}
\usepackage{ltablex}
\usepackage{url}
\usepackage[colorlinks=true, linkcolor=blue, citecolor=blue, urlcolor=blue]{hyperref}
\title{Introducing the United States Secretary of State (USSOS) Speech Dataset} 
\author{Benjamin Yoel\thanks{Postdoctoral Fellow, Stanford University, benyoel@stanford.edu} \& Jay Revolinsky\thanks{PhD Student, Computer Science, Michigan State University, revolins@msu.edu}}
\date{\today}

\begin{document}
\maketitle
\thispagestyle{empty}

\section*{Abstract}

\begin{singlespace}

\noindent A large number of theories in international relations attribute importance to the rhetoric of leading foreign policy actors. Moreover, in recent years, there has been a significant growth in interest for producing machine-readable text of remarks from key foreign policy actors. Yet, researchers have, to date, focused primarily on the rhetoric of executive leaders and members of legislatures. To understand the construction of foreign policy in the public arena and dynamics in it, especially in the United States, we need a collection of fine-grained textual representation of the remarks of US Secretaries of State. An especially promising use for such a corpus of text is to measure attention areas in US foreign policy and its implication for the US's policies, including in foreign aid. In this note, we introduce the United States Secretary of State Dataset, a corpus of 17,700 remarks delivered by US Secretaries of State from 1997 to 2026. Using this dataset, we demonstrate that Secretaries of State frequently make public remarks, the likes of which can have substantively meaningful implications. \href{https://github.com/revolins/sos-corpus}{Dataset Available Here.}

\end{singlespace}
\pagebreak 
\setcounter{page}{1}

\section*{Introduction}

On February 14, 2026, during his speech at the Munich Security Conference (MSC), U.S. Secretary of State, Marco Rubio, declared that, \textit{``We are part of one civilization – Western civilization.  We are bound to one another by the deepest bonds that nations could share, forged by centuries of shared history, Christian faith, culture, heritage, language, ancestry, and the sacrifices our forefathers made together for the common civilization to which we have fallen heir''} \citep{rubio2026munich}. Similarly, in July 2025, Secretary Rubio sent a diplomatic cable to US foreign service officers and diplomats abroad ordering them to refrain from commenting on elections abroad. In particular, he ordered them not to comment on their free and fair nature, and that if any comment were made, it would simply be to congratulate the winner \citep{lee2025trump}. The MSC speech and diplomatic cable both reflected a notable shift in the US's foreign policy and foreign policy rhetoric since the return of President Trump to the White House in early 2025, one that moved away from a longstanding focus on democracy through promotion of free and fair elections around the world \citep{bubeck2025}. 

Scholarly interest in understanding the foreign policy positions of states in the international system using text has risen significantly in recent years \citep[e.g.,][]{kentikelenis2021legitimacy, boucher2019tariff, turco2024speaking, sakamoto2023threat, bethke2024language, peez2026using, tokhi2025far, terman2023geopolitics}. This explosion has prompted scholars to also assess the implications of such rhetoric from foreign policy actors for public opinion and policymaking \citep[e.g.,][]{baum2004going, carter2020diversionary, gibson2025unpacking, peez2026using}. This increased interest is largely attributed to the recent proliferation of text-as-data methods, which enabled scholars to digitize and convert large corpora of text into machine-readable data \citep{grimmer2022text}. 

That said, while there has been an uptick in foreign policy rhetoric, we know relatively little about the rhetoric of key foreign policy actors in the international system, both in terms of its determinants and its effects. Most existing studies focus on the foreign policy rhetoric of executive leaders \citep[e.g.,][]{boucher2019tariff, kentikelenis2021legitimacy} or senior politicians in legislatures such as the US Congress \citep[e.g.,][]{bubeck2025}. In contrast to these set of actors, who must address foreign and domestic politics, foreign ministers are tasked with \textit{solely} focusing on crafting and executing foreign policy. Among existing research that does pay attention to foreign ministers and their political behavior, most tend to focus on the causes and consequences of these actors' visits to other countries \citep[see, e.g.,][]{lebovic2016diplomatic, malis2021state, mcmanus2025sending, mcmanus2017statements}. While important, visits to other countries only capture a small portion of their behavior and responsibilities. Despite the fact that foreign ministers often deliver public remarks, we know surprisingly little about what they say. A promising application for such data would be to capture the extent to which foreign policy actors speak about different policies and why they do so. For instance, when and why do foreign ministers speak about foreign aid?

In this paper, we present the USSOS Dataset, an original dataset of public communication by US Secretaries of State from 1997 to 2026, consisting of written and oral remarks. The unit of analysis is the individual remark made by the Secretary of State, which can be examined using both quantitative and qualitative text analysis. In short, the dataset includes 17,700 remarks made by Secretaries of State over this time period, covering six presidential administrations and nine Secretaries of State. Prior scholarship has highlighted the importance of public rhetoric for private diplomacy \citep{myrick2022making}.

Scholars may use the data to, for instance, examine longstanding debates about the role of norms in international politics. In this respect, research on the strength and evolution of the liberal international order could examine whether US Secretary of State rhetoric emphasizes democracy promotion or great power competition over time \citep[e.g.,][]{ikenberry2011liberal, deudney1999nature}, particularly given concerns of the eroding commitment of liberal democratic powers to democracy promotion and democratic backsliding within these countries \citep{hyde2025unconstrained}, as well as reduced commitment to international organizations \citep[e.g.,][]{tallberg2025democracy}. Relatedly, researchers can use the corpus of remarks to reexamine theories of credibility in international relations. Scholars may use the data to assess whether changes in rhetorical sentiment toward specific states or issue areas precede shifts in US policy behavior, thereby contributing to debates over whether public communication conveys meaningful information or is simply ``cheap talk'' \citep[e.g.,][]{fearon1994signaling, tomz2007domestic, tingley2011can, kertzer2020price}. Additionally, scholars can build on the existing literature on costly signals \citep[e.g.,][]{mcmanus2017statements, myrick2026today}, and examine whether the discourse of the US Secretary of State is viewed as costly to other countries on equal terms with international discourse from executive leaders. 

We first demonstrate descriptive data from the public communication of US Secretaries of State, showcasing trends in the extent of rhetoric by Secretaries across presidential administrations, including the timing of these speeches with the countries and international organizations they are most likely to reference. We demonstrate that our measure of US Secretary of State stance toward other countries is correlated with existing measures of alignment with the United States, particularly the UN General Assembly geopolitical alignment measure \citep{bailey2017estimating}. We then fit a structural topical model to the full corpus of remarks to better understand what topics Secretaries talk about. We also disaggregate this by party, and observe that certain topics are, as expected, concentrated among Secretaries from specific parties. For instance, Secretaries in Republican administrations are more likely to talk about the Israeli-Palestinian conflict, Iran and nuclear weapons, as well as Iraq. In contrast, Secretaries in Democratic administrations are more likely to stress the importance of climate change, Russia and Ukraine, and economic development. 

And lastly, we use these data to better understand public discourse from key foreign policy actors on foreign aid over time and attention to foreign aid across party lines in the United States. Building on the extensive literature on the political economy of foreign aid \citep{dietrich2015foreign, dietrich2020text, tokhi2025far, therien2000political, dreher2018apples, allen2018donor, greene2024donor, scott2011sponsoring}, we examine attention to foreign aid in the public communication of Secretaries of State. In particular, we build on and extend the work of \cite{dietrich2020text}. We find that, on average Democratic and Republican Secretaries of State do not differ significantly in how often they discuss foreign aid, in general. That said, Secretaries of State in Democratic administrations are significantly more likely to talk about aid in a positive manner than Republicans, and the net stance is lower for Republicans. 

We make several contributions to the study of international relations and US foreign policy. First, we introduce a novel corpus of Secretary of State rhetoric spanning much of the post-Cold War period, contributing to a growing literature on political speech in international politics. Second, we shift attention to a central but understudied foreign policy actor. Unlike sitting Presidents and Members of Congress, the Secretary of State is tasked primarily with formulating and communicating US foreign policy, making their rhetoric especially informative of US foreign policy. Third, the dataset captures communication across a wide range of contexts, from major speeches to routine remarks, delivered both domestically and abroad. This allows scholars to analyze variation in foreign policy attention across settings and audiences, including differences between domestic and international messaging.

This paper is structured as follows. We first present the United States Secretary of State Speech Corpus. We then introduce descriptive statistics of the corpus using a structural topic model, enabling us to illustrate the rich textual resource that this dataset offers to scholars of both US foreign policy and international relations. Then, to further illustrate the utility of the dataset, we develop a measure of attention to other countries and demonstrate that it can improve existing understandings of US foreign policy decision-making. 

\section*{The US Secretary of State Speech Corpus}

The US Secretary of State (USSOS) Speech Corpus is a collection of public communication by the US Secretary of State from 1997 to 2026. The dataset is intended to capture the \textit{public} remarks of US Secretaries of State and US foreign policy throughout the post-Cold War era. Overall, the corpus of public communication contains approximately 17,700 remarks, ranging from major speeches, press statements, testimonies, and press briefings. In this way, the corpus offers researchers data on different types of remarks by Secretaries. Understanding the discourse of senior US foreign policy officials is important since the United States has been \textit{``crucial to financing, coordinating, or mobilizing responses to major international problems or crises''} \citep[p.S72]{hyde2025unconstrained}. We focus on the post-Cold War era for two main reasons. First, the United States has played a particularly outsized role in shaping international politics during this period \citep[e.g.,][]{cooley2020exit}. Second, this focus is limited by data availability of the public discourse of US Secretaries of State .

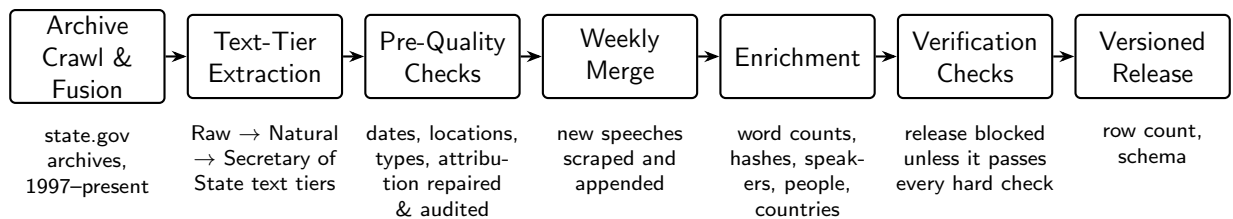
\begin{figure*}[t]
\centering
\begin{tikzpicture}[
  font=\footnotesize\sffamily,
  stage/.style={
    draw=black,
    fill=white,
    rounded corners=2pt,
    minimum height=11mm,
    text width=19mm,
    align=center,
    inner sep=2pt,
    thick
  },
  gate/.style={
    draw=black,
    fill=white,
    rounded corners=2pt,
    minimum height=11mm,
    text width=19mm,
    align=center,
    inner sep=2pt,
    thick
  },
  note/.style={
    font=\scriptsize\sffamily,
    text=black,
    text width=21mm,
    align=center
  },
  arr/.style={
    -{Stealth[length=2mm]},
    thick,
    black
  }
]

\node[stage] (crawl)   {Archive Crawl \& Fusion};
\node[stage, right=2.8mm of crawl] (extract) {Text-Tier Extraction};
\node[stage, right=2.8mm of extract] (final) {Pre-Quality Checks};
\node[stage, right=2.8mm of final] (weekly) {Weekly Merge};
\node[stage, right=2.8mm of weekly] (enrich) {Enrichment};
\node[gate, right=2.8mm of enrich] (verify) {Verification Checks};
\node[stage, right=2.8mm of verify] (release) {Versioned Release};

\draw[arr] (crawl) -- (extract);
\draw[arr] (extract) -- (final);
\draw[arr] (final) -- (weekly);
\draw[arr] (weekly) -- (enrich);
\draw[arr] (enrich) -- (verify);
\draw[arr] (verify) -- (release);

\node[note, below=2mm of crawl]
  {state.gov archives, 1997--present};

\node[note, below=2mm of extract]
  {Raw $\rightarrow$ Natural $\rightarrow$ Secretary of State text tiers};

\node[note, below=2mm of final]
  {dates, locations, types, attribution repaired \& audited};

\node[note, below=2mm of weekly]
  {new speeches scraped and appended};

\node[note, below=2mm of enrich]
  {word counts, hashes, speakers, people, countries};

\node[note, below=2mm of verify]
  {release blocked unless it passes every hard check};

\node[note, below=2mm of release]
  {row count, schema};

\end{tikzpicture}
\caption{Secretary of State Speech Dataset Production pipeline.}
\label{fig:pipeline}
\end{figure*}

The data are available in Comma-separated values (CSV) form, R and Python packages, and a dedicated website hosted on GitHub. We present each document in plain text format. Each row in the dataset represents one remark by the US Secretary of State, including a major address, press statement, or a testimony to Congress. We include the raw text, that covers both Secretaries of State and other speakers such as foreign dignitaries, the cleaned and transformed text, the proportion of remarks delivered by the Secretary, and text that only includes the words explicitly delivered by the Secretary. We conduct a host of verification checks, summarized as a seven-stage pipeline in Figure~\ref{fig:pipeline}. The pipeline begins with scraping from the Digital Archives of the Department of State, then involves extracting the raw text from the documents, pulling important metadata such as dates, automatic scraping of new public statements, enrichment, a verification check involving 83 different procedures to ensure the data is robust and accurately coded, and then released onto GitHub weekly. The data generating process is intended to be replicable, transparent, and extend into the future automatically.\footnote{We apply Claude Opus and Claude Sonnet to assist with the automated management process.} We document the full process in more detail in the Appendix.

On average, there are 1000 words in each set of remarks, though this varies across Secretaries and the type of remark. Perhaps unsurprisingly, press statements issued by Secretaries of State tend to be, on average, significantly shorter than major addresses and press conferences. One of the main advantages of the USSOS Speech Dataset is that it covers three decades across the post-Cold War era, spanning Republican and Democratic administrations and shifting foreign policy policy priorities. Additionally, it captures remarks delivered in the United States and abroad. We scraped the remarks of Secretaries from the Archives of the Department of State website, which digitized the remarks of Secretaries of State from Madeleine Albright until Marco Rubio (January 1997 through September 2026).

\section*{The Dataset At A Glance}

\begin{figure}[H]
    \centering
    \includegraphics[width=\textwidth]{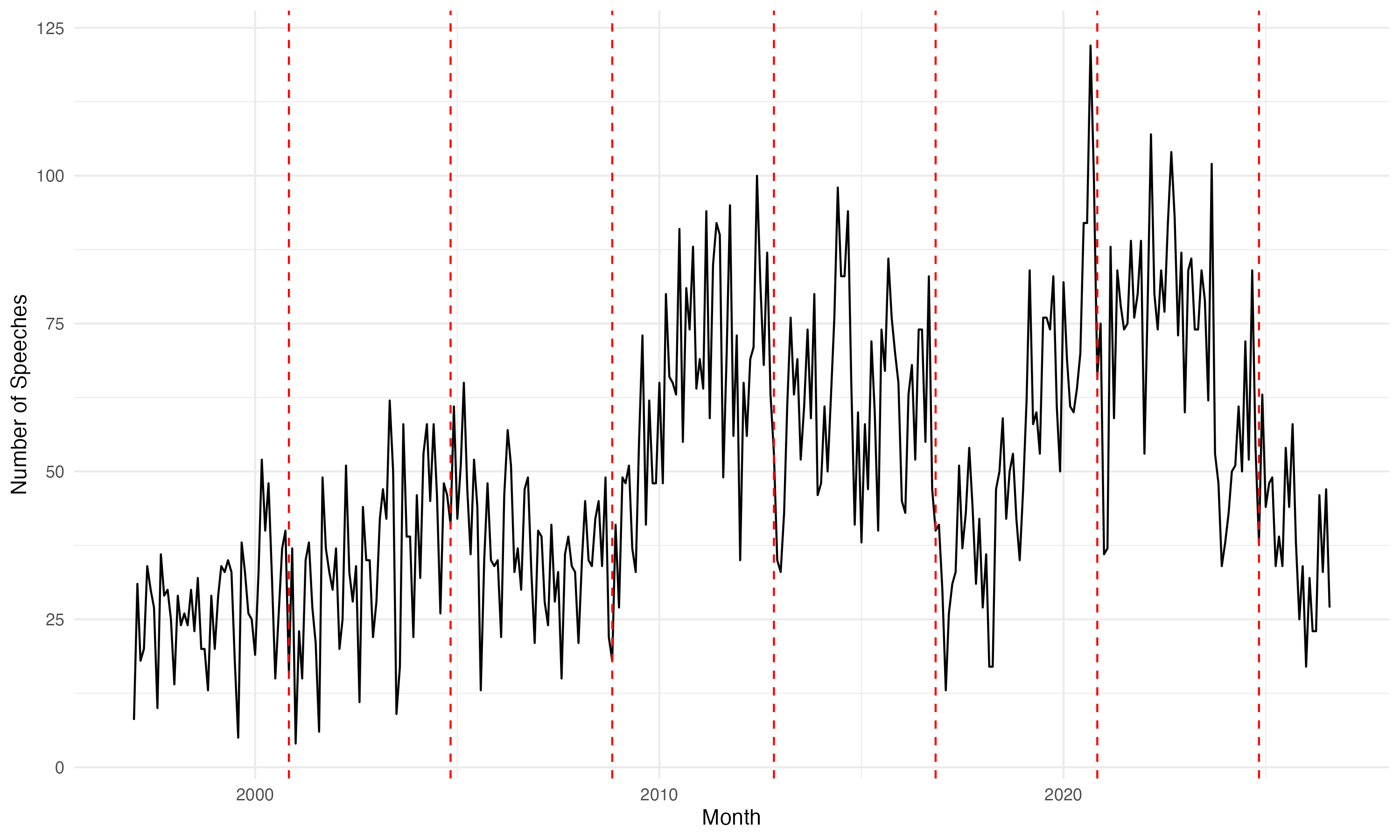}
    \caption{Link Between Remarks by Secretaries of State and Presidential Elections}
    \label{fig:preselections}
\end{figure}

To assess the extent on which US Secretaries of State engage in public communication, Figure~\ref{fig:preselections} depicts the number of remarks each Secretary of State delivered in each month for the period between January 1997 to September 2026. The figure includes dashed red lines for US presidential election-months. In short, we observe that Secretaries of State frequently deliver public remarks. Temporally, there is interesting variation across the corpus. We observe a sharp increase in remarks made by Secretaries over time, particularly compared to earlier ones. There is a substantial uptick in remarks delivered by Secretaries during Barack Obama's presidency, particularly during Clinton's tenure as Secretary of State (2009-2013). 

We then observe a notable decline in the number of remarks delivered during the first term of President Donald J. Trump, particularly for his first Secretary of State, Rex Tillerson (2017-2018), a former businessman and political outsider. Then, we observe a return to a large number of remarks being delivered by Secretary Mike Pompeo, a former United States Congressman, from mid-2018 until the end of his term and the end of the first administration of President Trump (January 2021). We then observe an uptick in speeches for President Joe Biden's Secretary, Antony Blinken (2021-2025) and then a decline in remarks around 2023 with an uptick in those remarks in the latter half of 2024. Finally, under President Trump's second term in office, Marco Rubio has delivered similar numbers of speeches to Secretary Tillerson, but considerably fewer than most other recent Secretaries of State in our corpus.

In addition to temporal variation in the \textit{number} of public remarks, Figure~\ref{fig:preselections} suggests that there is some variation in the timing of these remarks by US Secretaries of State. This is particularly apparent for some Secretaries in the lead-up to presidential election-months. For some Secretaries, we observe an uptick in the lead-up to US presidential election-months, particularly Condoleezza Rice and Mike Pompeo. This adds nuance to prior work that suggests that US foreign policy actors shift their attention away from foreign policy areas, particularly elections, abroad and focus on the domestic sphere \citep{bubeck2025}. While that work focuses on Presidents and Members of Congress, our descriptive data suggest that Secretaries of State can potentially step in and help presidents gain foreign policy victories.

As noted above, during the process of converting the documents into clean text, we collected important metadata, including the location of remarks, length of remarks, as well as speakers outside the Secretary of State. In this way, scholars can use the data to, for instance, examine differences in rhetoric on visits compared to remarks in the US. After collecting the data, we pre-processed the data, including stemming, lemmatizing, and removing stopwords, building on existing work in machine learning \citep{grimmer2022text}.

By tracking the public discourse of the US Secretary of State across three decades and different presidential administrations during the post-Cold War period (1997-2026), we can construct a measure of attention to different policy issues by US Secretaries of State, including foreign aid, nuclear weapons, and climate change. While existing scholarship has acknowledged the important role of the US Secretary of State, much of this research focuses on their visits to other countries \citep{lebovic2016diplomatic, malis2021state, mcmanus2025sending, mcmanus2017statements}. While important as prior research finds, particularly for shaping public opinion in other countries and increasing the odds of foreign leader survival, visits by Secretaries of State only make up a fraction of their activities. 

\section*{Descriptive Statistics}
We now proceed to highlight several contributions that the USSOS Speech Dataset offers to scholars of US foreign policy and international relations. First, we shed light on \textit{who} and \textit{where} Secretaries of State talk. Second, we examine the topics that US Secretaries of State opt to focus on in their remarks using advances in unsupervised machine learning \citep{roberts2019stm, grimmer2022text} to better understand the policy priorities of Secretaries of State across US presidential administrations.

In doing so, we demonstrate the significant heterogeneity in countries and international organizations discussed, as well as the topics and issue areas Secretaries of State focus on over time. We then showcase the utility of the dataset of remarks through an application to the literature on foreign aid, specifically the strategic allocation of foreign aid. These analyses demonstrate the heterogeneity in topics brought up in the remarks of US Secretaries of State and reveal a nuanced understanding of US foreign policy attention. These data, as such, can be combined to understand foreign policy priorities and preferences of the United States, both within and across party lines, including the strategic nature of US foreign policymaking.

\subsection*{Where and Who Secretaries of State Talk About}

We now shed light on the \textit{locations} in which Secretaries of States deliver their remarks as well as \textit{which} countries and international organizations they are especially likely to reference in their public communication. In essence, examining \textit{where} US Secretaries of State deliver their remarks speaks to the importance of diplomatic visits by leading officials highlighted by others \citep{malis2021state, mcmanus2025sending, lebovic2016diplomatic}, which scholars emphasize as being important signals of resolve \citep{mcmanus2017statements}. Scholars could use these data to examine variation in policy priorities and focus areas depending on the sentiment of remarks delivered to domestic compared to international audiences. Relatedly, zooming in on the countries and organizations Secretaries reference can inform us about whether they focus their attention on the US's allies and strengthening alliances, or on attempting to discredit or shame their adversaries for human rights violations \citep{terman2023geopolitics, kentikelenis2021legitimacy, bailey2017estimating}. 

\begin{figure}[H]
    \centering
    \includegraphics[width=0.9\textwidth]{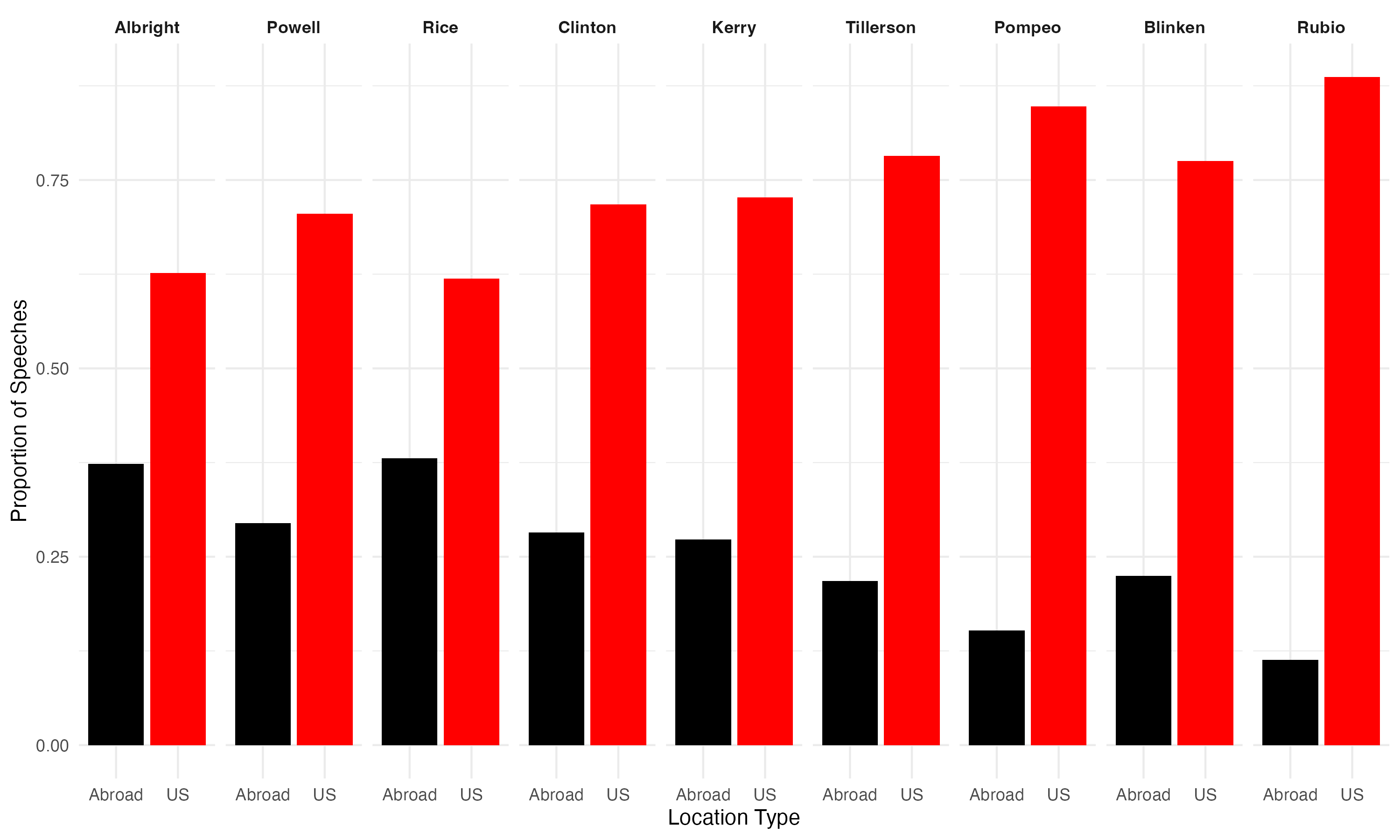}
    \caption{Location of Remarks by Secretaries of State}
    \label{fig:soslocations}
\end{figure}

First, we examine \textit{where} Secretaries of State are most likely to deliver public communication over time. Here, we focus on the degree to which they speak domestically, namely within the United States, and the degree to which they speak while on visits abroad. Figure~\ref{fig:soslocations} reveals, perhaps unsurprisingly, that Secretaries mostly deliver their public communication in the United States. However, a small but significant proportion of remarks are delivered during their visits abroad, further highlighting the importance of public diplomacy. In particular, it suggests, as in Figure~\ref{fig:preselections}, that Secretaries are consistently delivering more remarks, but that they are increasingly doing so at home, though especially in Washington, DC.

\begin{figure}[H]
    \centering
    \includegraphics[width=0.85\textwidth]{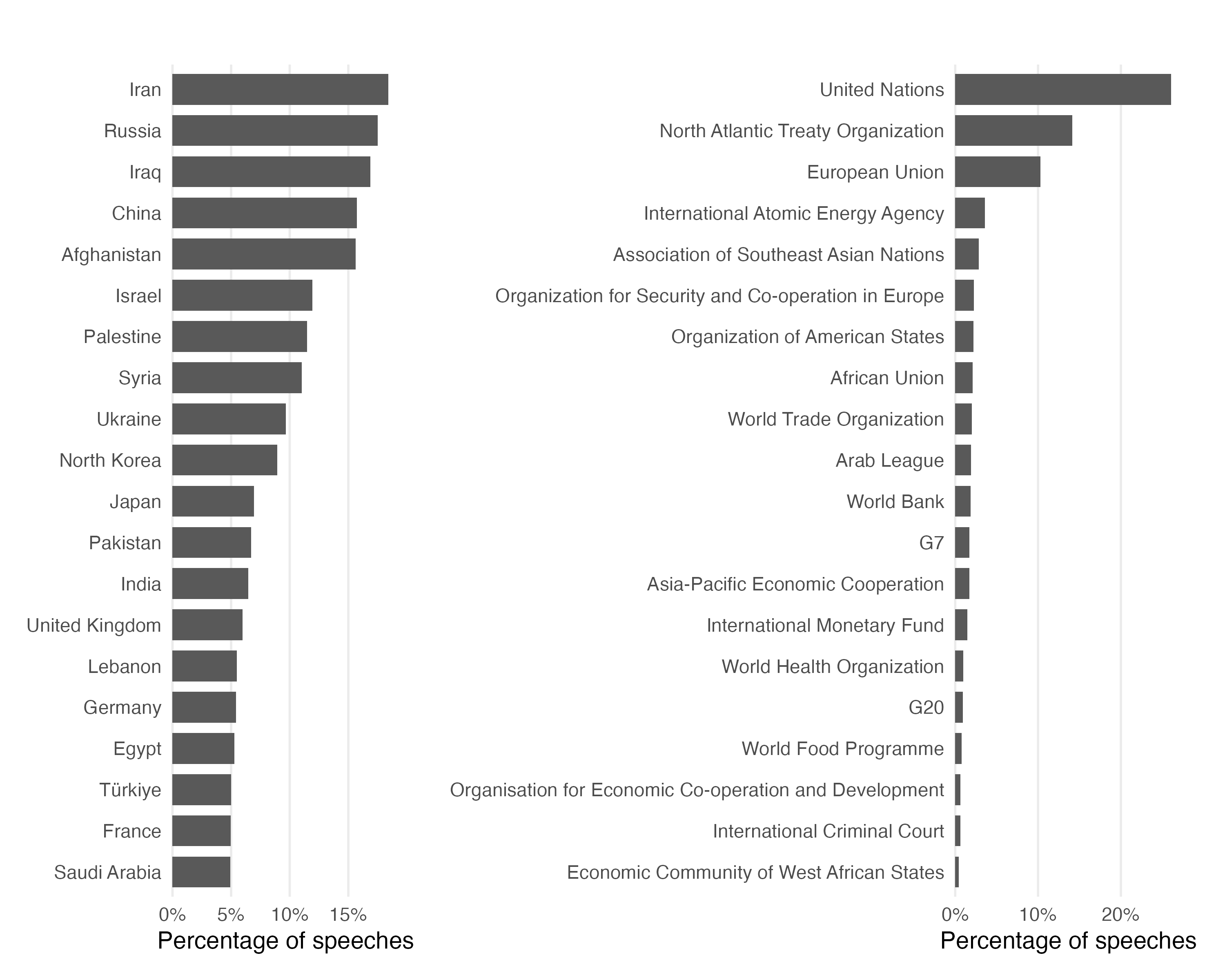}
    \caption{Who Secretaries of States Reference in their Remarks}
    \label{fig:whowhere}
\end{figure}

Now, we turn to examine \textit{who} US Secretaries of State are most likely to talk about in their public communication. More specifically, we focus primarily on governmental actors Secretaries of State talk about, namely other \textit{countries} Secretaries focus on, as well as global and regional international organizations. We identify countries and international organizations by references to their names (e.g., United Arab Emirates) or acronyms (e.g., UAE) using the \textit{countrycode} package in R, on top of names of foreign capitals. In this way, we can capture references to historical names of countries as well as their current name, such as Yugoslavia and Serbia. We observe that all UN Member States get mentioned at least once by a Secretary of State, as captured within the current SOS Speech Dataset. That said, some Secretaries of State do not mention a handful of countries, particularly smaller states in the Pacific Islands.\footnote{For the full list and statistics, see Figure~\ref{fig:secstatecountries} of the proportion of speeches in which different countries get mentioned and Table~\ref{tab:countriesmentioned} for countries not mentioned by a given Secretary of State.} Figure~\ref{fig:whowhere} plots the top twenty countries and international organizations referenced across all Secretaries. The figure reveals that Secretaries of State tend to focus their public discourse especially on the US's adversaries, but also some allied countries.

Across administrations, Iran, Russia, and China receive substantial attention in the public discourse of US Secretaries of State. During the first half of the dataset's coverage (1997-2011, in particular), Secretaries tend to focus especially on Iraq and Afghanistan, unsurprising given that the US sought to impose regime change onto these countries \citep{downes2013forced}. That said, Secretaries also reference US allies, such as Israel, Japan, the United Kingdom, and France. Meanwhile, with respect to international organizations, other examples indicates the United Nations receives the most attention, followed by NATO and the European Union, a trend that is consistent across Secretaries. In the appendix, we show that Secretaries of State reference most countries at least once during their tenure (Figure~\ref{fig:secstatecountries}). Moreover, there is some variation in the frequency of mentions of different countries, though Democratic and Republican Secretaries of State are likely to mention a very similar set of countries (Figure~\ref{fig:countriesparty}). 
On top of that, we include Figure~\ref{fig:across} as an example of how attention towards Ukraine shifts in the later half of the dataset's coverage.

Additionally, Figure~\ref{fig:network} provides a preview for future network analysis, where scholars could investigate communication patterns using the USSOS Speech Dataset. In doing so, the figures depicts co-presence and co-mentions of separate speakers with explicit structure. Both forms of evidence, shown in Figure~\ref{fig:across} and  Figure~\ref{fig:network}, indicate some partisan differences in attention to different countries in the public communication of the US Secretary of State. However, our ongoing analysis indicates these differences are concentrated among a small number of Secretaries and largely coincide with periods of active military intervention (e.g., the invasion of Iraq) or diplomatic engagement (e.g., the Iran nuclear deal negotiations) rather than a stable partisan pattern in foreign policy attention, in line with the logic of a ``bipartisan consensus'' in foreign policymaking \citep[e.g.,][]{friedman2024myth, tama2024bipartisanship}. That said, the effects of discourse and political attention on foreign relations represents an important area for future research, especially in a direction that could be systematically explored.

\subsection*{Validation Against Existing Measures}

To validate whether our measure reliably captures US foreign policy stances toward other countries over time, we examine the relationship between our measure of Secretary of State attention respective to other measures used to capture alignment with the United States. More specifically, we compare our Secretary of State's stance measure with the measure of voting alignment at the UN General Assembly (UNGA) from \cite{bailey2017estimating}. The UNGA voting measure has been used extensively in existing scholarship on both US foreign policy and diplomacy, more broadly. In short, it offers a range of advantages, particularly that it captures the extent to which countries are aligned with each other over time. That said, it is restricted to the \textit{country-year} level, and focuses only on the UN General Assembly. While important, this constitutes only one forum of diplomacy in the international system. We suggest that a focus on the Secretary of State can help shed light on the broader sentiment and stance of the United States vis-a-vis other countries. 

We use the discourse of US Secretaries of State to separate sentiment and stance as independent measures. As recent work in political methodology notes, sentiment and stance are unique to one another \citep{griswold2026stay, bestvater2023sentiment}. For instance, \cite{bestvater2023sentiment} argue, \textit{``sentiment and stance are distinct concepts that, although related, cannot necessarily be treated as interchangeable''} (p.253). 

More specifically, we build on \cite{griswold2026stay} and use large language models (LLMs) to code Secretaries' stance and sentiment. For our purposes, sentiment refers to the emotion of remarks by Secretaries about other countries, whereas stance captures political motivation, focusing primarily on targeted intent towards another country. The codebook, as shown in Section~\ref{subsec:llm-cb-country}, ensures that the LLM clearly separates sentiment and stance. We construct a score of stance for each country Secretaries reference with three possible values (-1,0,+1), where -1 is negative, 0 is neutral, and +1 is positive. For example, a sentence condemning a warlord for blocking aid disbursement is coded as negative in sentiment, but indicates a supportive stance towards the citizens of the affected country.

In contrast to existing measures, our stance measure can be used at a range of units of analysis, such as the country-month. If our measure accurately captures the US's policies and stance on the international stage vis-a-vis other countries, we should observe a strong correlation between our measure and the UN General Assembly voting alignment measure. To this end, we aggregate up alignment to the country-year level, such that each country-year gets its own correlation score, which we then average to get the overall correlation. We follow existing studies and measure alignment with the US as the difference between a country's ideal point score and that of the United States, such that countries with higher scores are \textit{least aligned} with the US, and those countries with lower scores are \textit{most aligned}. 

\begin{figure}[H]
    \centering
    \includegraphics[width=0.73\textwidth]{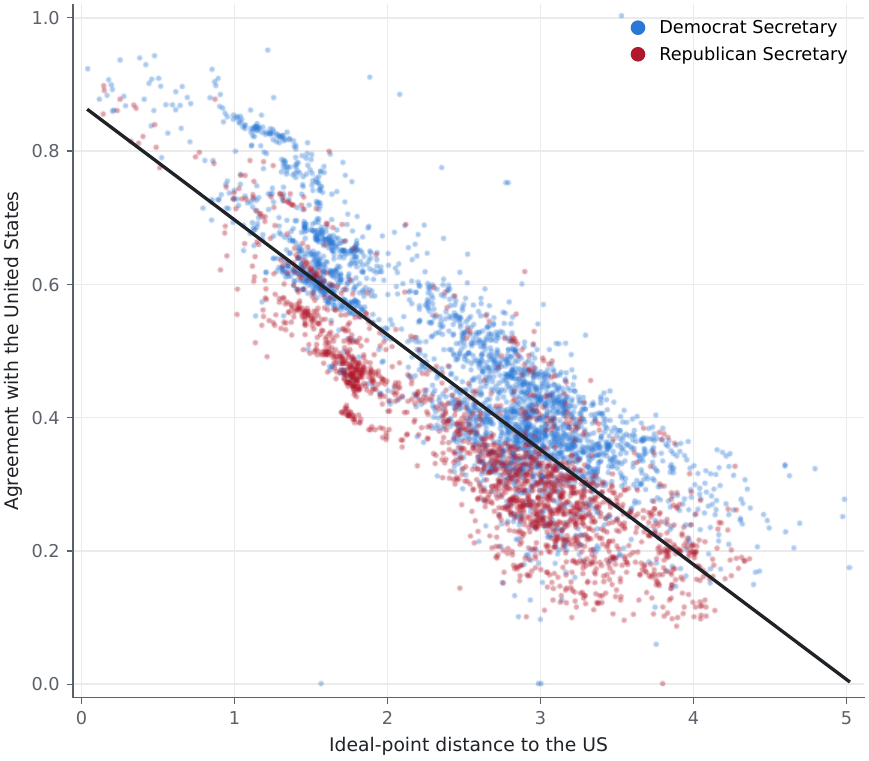}
    \caption{Correlation Between Secretary of State Measure and UNGA Voting Alignment}
    \label{fig:ungavoting}
\end{figure}

As shown in Figure~\ref{fig:ungavoting}, the correlation ($\rho$) between Secretary of State speech corpora and UN voting alignment remain remarkably stable across US Secretaries of State, holding at a consistent pooled correlation of $0.82$. That said, there is some fluctuation in the correlation between our measure and the UNGA voting alignment measure, with the correlation being lower for the years of the Bush administration (2001-2009). Overall, the measure is meaningfully correlated yet distinct from voting alignment at the UN General Assembly. 

Rather than evaluating how countries vote, this suggests that our measure reliably tracks the administration's broader strategic orientation and foreign policy positions vis-a-vis other countries. While we cannot, and do not claim that our measure captures geopolitical alignment with other countries, future work could similarly collect data of foreign minister and executive leader rhetoric of other great powers, such as Russia and China to construct a measure of geopolitical alignment with the United States. Within the Appendix, Figure~\ref{fig:unga_by_sec} details a per-country and per-secretary breakdown of Figure~\ref{fig:ungavoting}. Together, both figures indicate strong similarities between our measure and the UNGA alignment measure, with a mean correlation of $\rho = 0.93$ between each Secretary of State from 1997 to 2026. Within Appendix~\ref{fig:temporal_sos}, we further analyze the per-secretary correlation over time, demonstrating alignment between our stance measure and the UNGA voting alignment measure.

\subsection*{The Topics Secretaries Focus On}

Next, we shed light on the precise substantive topics Secretaries of State focus on. To this end, we fit a structural topic model (STM), an unsupervised form of machine learning that enables the model to learn topics ``along the way,'' and one that has been frequently used to better understand trends in both public and private discourse and documents with large samples \citep[e.g.,][]{roberts2019stm, grimmer2022text}. 

We fit the STM onto the full corpus of public remarks of the US Secretary of State, examining the \textit{topics} that dominate the remarks of Secretaries of State across administrations.  For the STM analyses, we decide to keep ``diplomatic speak'' in our models, namely rhetoric that is merely symbolic because the rhetoric is often used in the practice of diplomacy and public communication across countries. Moreover, as \cite{gibson2025unpacking} argues, \textit{``detection of a shift in the way one state refers to the leader of another state can be a useful indicator of upcoming policy shifts''} (p.28). We identified words that constitute ``diplospeak'' and pleasantries by drawing on the literature on public diplomacy, as well as deductive interpretation. Words that constitute ``diplospeak'' include candid and exchange, while pleasantries include phrases such as ``thank,'' alongside individual names of Secretaries of State and the president. 

To determine the appropriate number of topics for our STM, we rely on the searchK function in the \textit{stm} package \citep{roberts2019stm}. We estimated structural topic models which vary between 5 topics and as many as 50 topics across the entire corpus of remarks. We find that the ideal number of topics is 20, as determined by four diagnostic values (see Figure~\ref{fig:diagnostics}). The STM model was selected on 20 topics since it balances semantic coherence, all while minimizing loss of distinctiveness from the remaining three metrics. 

\begin{figure}[H]
    \centering
    \includegraphics[width=\textwidth]{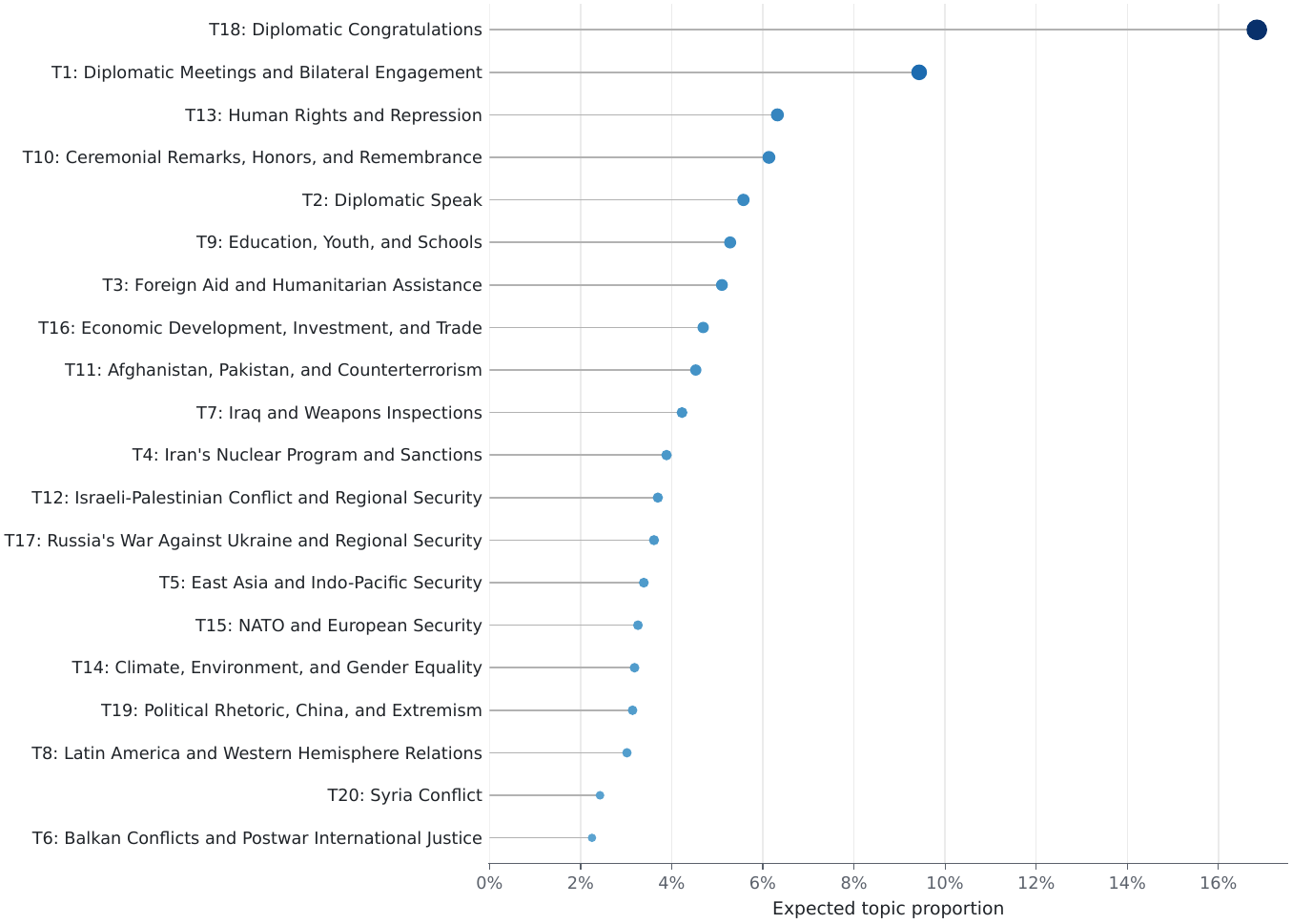}
    \caption{Most Common Topics in the Remarks of US Secretaries of State}
    \label{fig:stm_main}
\end{figure}

Figure~\ref{fig:stm_main} depicts the 20 topic clusters for the discourse of Secretaries of State. In short, it suggests that Secretaries of State across administrations talk about a range of topics. In particular, Secretaries of State engage extensively in ``diplomatic speak''. Beyond diplomatic speak (Topic 2), Secretaries of State appear to extensively draw attention to human rights and democracy (Topic 13), economic development (Topic 16), foreign aid (Topic 3), and climate change and women's rights (Topic 14). Moreover, Secretaries of State are especially likely to talk about international interests and promoting the United States's role in the world. In addition, we find that Secretaries of State are especially likely to reference presidents (a pattern observed across different topics), particularly the US President. The model also highlights the consistent presence of regional wars and crises, such as the Israeli-Palestinian conflict (Topic 12), the war in Ukraine (Topic 17), China and great power rivalry (Topic 5), the NATO alliance and Russia (Topic 15), Iraq (Topic 7), Afghanistan (Topic 11), and Iran (Topic 4). Accordingly, our STM suggests that Secretaries of State in the post-Cold War era talk about a range of policy areas in the foreign policy space. This, we argue, makes the data especially useful for scholars across different areas in international relations.

\begin{figure}[H]
    \centering
    \includegraphics[width=0.83\textwidth]{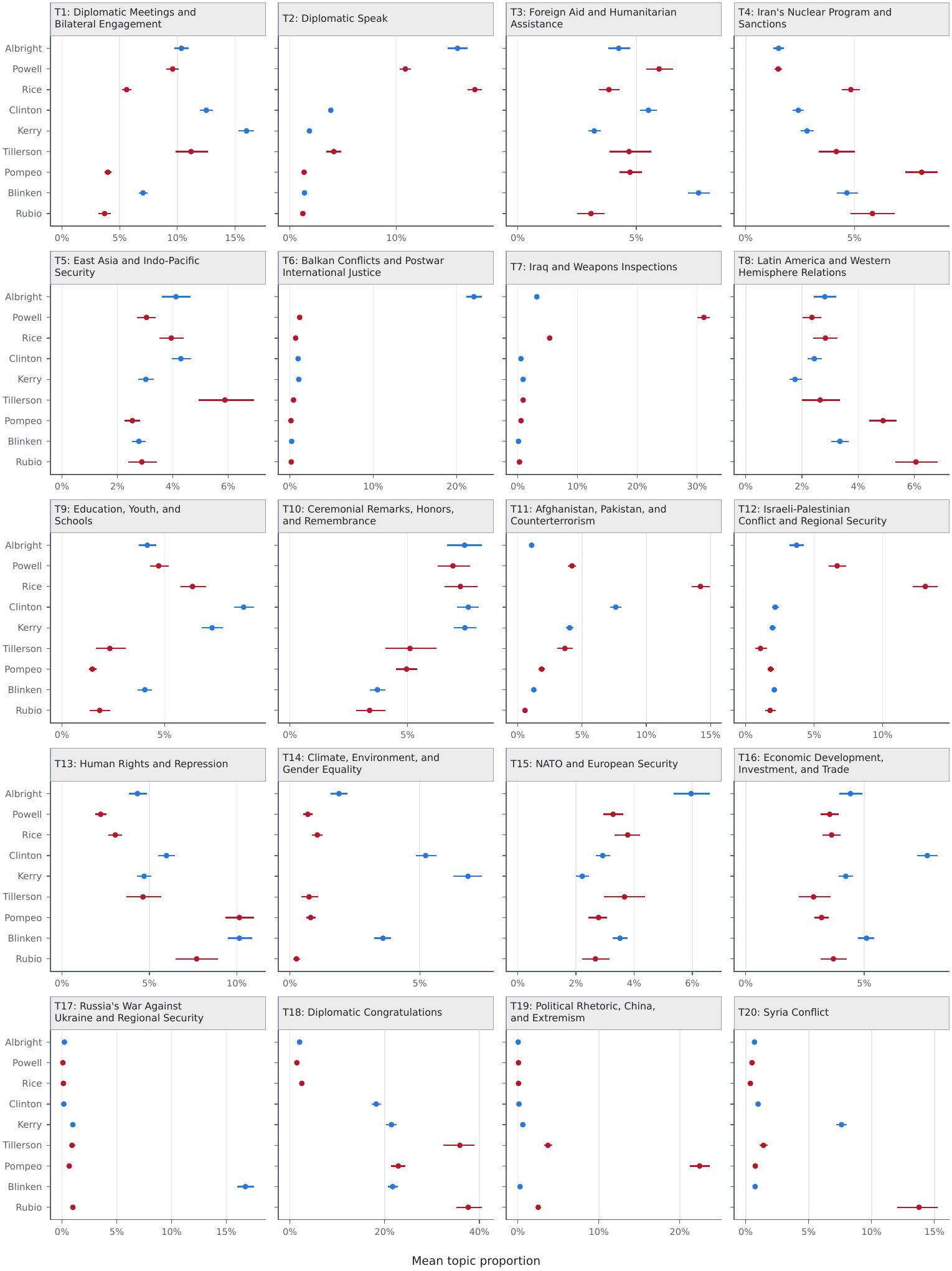}
    \caption{Most Common Topics for Each Secretary of State}
    \label{fig:stm}
\end{figure}

Beyond the most common topics, we also disaggregate by the topics that \textit{each} Secretary of State focuses on in Figure~\ref{fig:stm}. In which, we indicate the proportion each Secretary devotes to that topic, capturing their party by presenting Democrats in blue and Republicans in red.

 Perhaps unsurprisingly, Figure~\ref{fig:stm} shows only one or a handful of Secretaries focus on certain topics, such as the NATO intervention in Kosovo and the Bosnia War (Topic 6) the war in Iraq and concerns of weapons of mass destruction (Topic 7), as well as the War in Ukraine (Topic 17). On others, we observe at least some mention of topic across different Secretaries of State such as foreign aid and assistance (Topic 3), the Western Hemisphere and Latin America (Topic 8), NATO and threats to it (Topic 15), human rights (Topic 13), and economic development (Topic 16). Collectively, this suggests that, while some topics and policy areas are specifically mentioned only by certain Secretaries, there is significant overlap across US administrations. The plot also suggests shifting priorities of US administrations. For instance, attention to the Iranian nuclear issue has gone up over time (Topic 4).

Moreover, given the large literature on the partisan sources of foreign policy preferences \citep[e.g.,][]{aldrich2006foreign, milner2015sailing, kertzer2020price}, we expect foreign policy rhetoric to vary systematically across Republican and Democratic administrations. While foreign policy in the United States is often characterized by substantial elite consensus \citep{friedman2024myth, tama2024bipartisanship}, existing scholarship demonstrates that partisan identities and polarization shape the issues that political actors prioritize and the types of global cooperation they favor \citep{hyde2025unconstrained}. Thus, examining variation in Secretary of State rhetoric provides an opportunity to assess whether these partisan differences extend to the public communication of US foreign policy actors. 

\begin{figure}[H]
    \centering
    \includegraphics[width=\textwidth]{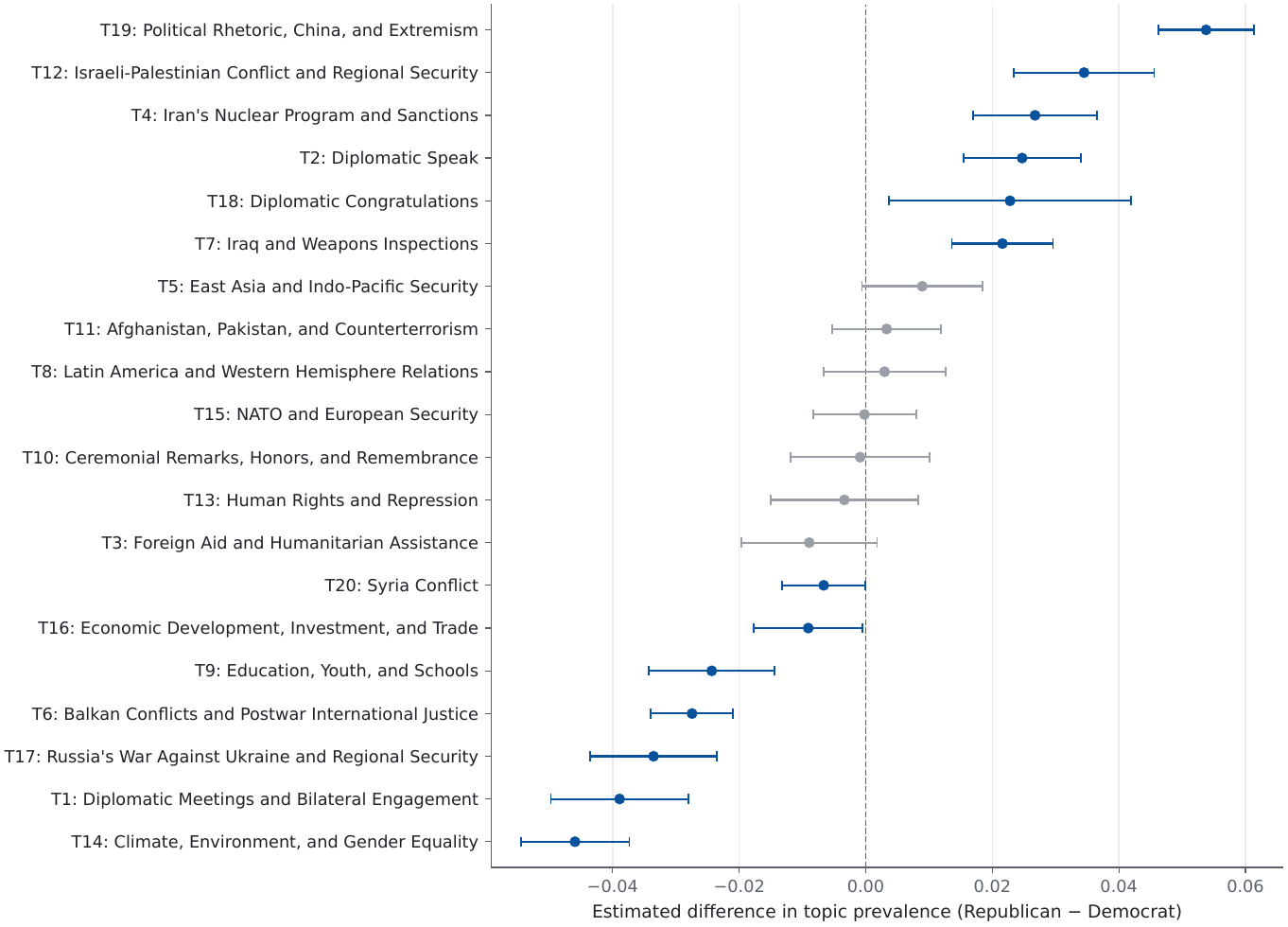}
    \caption{Topics of Secretary of State Speeches by Party}
    \label{fig:type}
\end{figure}

We observe some \textit{partisan} differences in the \textit{topics} mentioned by Secretaries of State. Figure~\ref{fig:type} depicts the estimated difference in topic prevalence of each of the 20 topics from the STM estimated above. The figure suggests that Secretaries of State in Democratic presidential administrations, relative to Republican Secretaries, are more likely to talk about: women's rights and climate change, the importance of the US as a leader of the international order, Ukraine and Russia, and economic development. In contrast, Republican Secretaries of State are especially likely to focus on: Israel and the Palestinians, Iran and nuclear weapons, diplomatic speak, and issues surrounding Iraq and Afghanistan, as well as China.

\section*{Application: Discourse on Foreign Aid}

We argue that topic proportions generated from the USSOS Speech Dataset can be informative of the US's foreign policy priorities. One area where this may be especially informative is in the realm of foreign aid (Topic 3 in our structural topic model).\footnote{The FREX from Topic 3 suggests the following words all constitute references to aid: \textit{aid, hiv, budget, usaid, sudan, request, fund, kenya, emergency, ethiopia, estimate, somalia, refugee, donor, humanitarian}.} 

An extensive amount of literature examines the causes and consequences of foreign aid flows from leading donors such as the United States and the European Union for a range of different political and economic outcomes \citep[e.g.,][]{scott2011sponsoring, finkel2007effects, dietrich2015foreign, knack2004does, milner2015sailing, lumsdaine1993moral}. In particular, existing work on the political economy of foreign aid highlights the importance of ideology in both political elite and party rhetoric on foreign aid as well as government \textit{spending} on foreign aid in democracies \citep{dietrich2020text, dreher2024aid, milner2015sailing, lumsdaine1993moral, therien2000political, allen2018donor, tokhi2025far, greene2024donor}. 

Here, we build directly on and extend the work of \cite{dietrich2020text}. These authors examine differences between political parties in their positions on foreign aid in party manifestos across 13 donor countries, including the United States. The authors find left-leaning political parties are more likely to speak about foreign aid versus right-leaning parties, \textit{and} that they are more likely to speak positively of aid than right-leaning parties.

We extend this argument, empirically examining it in the context of the rhetoric of US Secretaries of State. In doing so, we probe the extent to which they discuss foreign aid and their stance, particularly how this varies between Republican and Democratic administrations. We posit that examining this argument in the context of the rhetoric of US Secretaries of State is especially important given the critical role of the State Department in setting an administration's foreign policy priorities in the US. In our sample, we have four Democratic Secretaries and five Republican Secretaries, covering three decades (1997-2026).

We draw on the arguments advanced directly in \cite{dietrich2020text} to derive expectations. First, we examine the extent to which Secretaries of State across party lines reference foreign aid. In short, we expect \textit{Democratic} Secretaries to be more likely to reference foreign aid than Republicans, \textit{overall}. Second, we highlight the stance of such rhetoric on foreign aid and variation in ideology. In so doing, we expect Secretaries of State from the Democratic Party to be more likely to reference aid in a positive manner than Republican Secretaries of State. In contrast, we expect Republican Secretaries of State to be more likely to reference aid negatively than Democratic Secretaries of State. 

To test these expectations in the context of US Secretary of State rhetoric, we construct a measure of attention to foreign aid in Secretary of State discourse applying the topic scores from our initial structural topic model test. More specifically, we use Topic 3, which our STM identified as focusing on foreign aid. It is important to note that this methodology differs from \cite{dietrich2020text} in several ways. First, \cite{dietrich2020text} focus on \textit{manifestos} of political parties across established democracies, whereas we examine the rhetoric of foreign ministers during their times in political office, specifically the US Secretary of State. Second, \cite{dietrich2020text} measure attention to foreign aid as the \textit{``number of mentions of foreign aid in the texts of manifestos per 1,000 words''} (p.984), and rely on human coders to identify mentions of foreign aid in individual sentences. In contrast, we use topic proportions from the structural topic model in Figure~\ref{fig:stm_main}. In this way, our measure is constructed using an approach that allows the model to learn from the data. To account for the fact that references to aid in the rhetoric of Secretaries of State are not independent and identically distributed, we cluster by Secretary of State-year.

Within our test, we examine whether Republican Secretaries of State are more likely to speak negatively or positively about foreign aid than Democratic Secretaries of State. In this regard, we distinguish between negative and positive remarks about foreign aid. To capture this distinction empirically, we use large language models (LLMs), which have become increasingly used to analyze political texts, particularly sentiment \citep[e.g.,][]{ornstein2025train, benoit2025using, griswold2026stay}. More specifically, we make use of Qwen3.8-27B LLM~\citep{qwen38_27b} to label sentiment and stance. We provided the LLM labeler with a system prompt
(Section~\ref{subsec:llm-cb-country}), where the instructions
for coding each example treat sentiment as the emotional tone and stance as the political intent described within a given Secretary's utterance. In this way, we can measure whether a remark made by a Secretary referencing foreign aid is positive, negative, or neutral across both sentiment and stance. Our application is centered entirely around \textit{political stance}.

\begin{figure}[H]
    \centering
    \includegraphics[width=\textwidth]{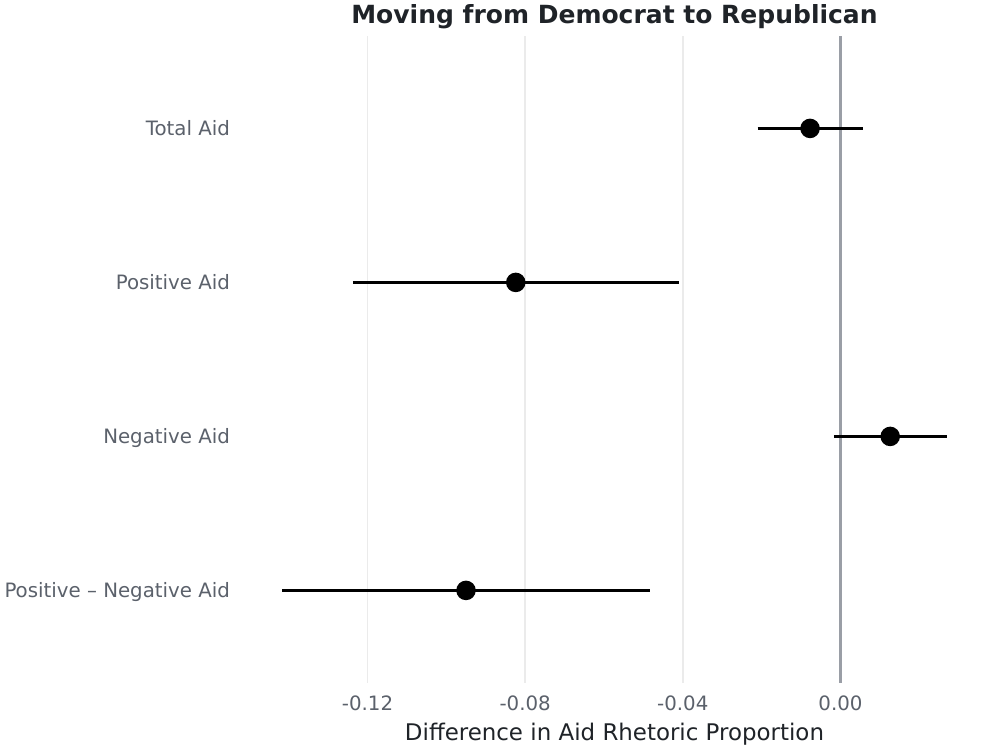}
    \caption{Differences in Foreign Aid Rhetoric By Party}
    \label{fig:foreignaiddms}
\end{figure}

Figure~\ref{fig:foreignaiddms} depicts the effects of party on references to our foreign aid topic proportion. We treat Republican Secretaries of State as being right-wing and Democrats as left-wing, consistent with \cite{dietrich2020text}. First, we seek to understand which Secretaries of State are most likely to mention foreign aid, in total. We find that there is no significant effect of ideology on the proportion of references to foreign aid in the discourse of US Secretaries of State. While Democratic Secretaries are slightly more likely than Republicans to talk about aid, in general, the differences are not significant. In other words, we do not find that Republican Secretaries are significantly less likely than Democratic Secretaries to reference foreign aid. This is consistent with the (null) finding from \cite{dietrich2020text} on the frequency of foreign aid discourse across partisan lines.

Next, we turn to examine variation in \textit{stance} in foreign aid discourse across party lines. In short, Figure~\ref{fig:foreignaiddms} indicates striking differences in stance  across party. We find that when Secretaries talk about foreign aid, on average, Democrats are more likely to speak positively about aid than Republicans, an effect that is significant and substantively meaningful ($p$ $<$ 0.001). When referencing foreign aid in their discourse, Democratic Secretaries use positive rhetoric at roughly 8 percent difference compared to Republican Secretaries. Meanwhile, we find weaker evidence that Republican Secretaries speak more negatively about foreign aid. This difference is small and only marginally significant ($p$ $<$ 0.08). As such, the partisan gap is driven mainly by Republicans being less positive rather than more negative. The net sentiment (the share positive of rhetoric minus the share of negative rhetoric) is, accordingly, lower for Republicans, by roughly 9.5 percentage points. In short, we find evidence within the USSOS dataset which is largely consistent with \cite{dietrich2020text}.

This partisan divide in rhetoric on foreign aid and the stance of Secretaries on aid is largely consistent across individual Secretaries of State, with Marco Rubio, President Donald Trump's current Secretary of State (2025-present), being an important exception (see Figure~\ref{fig:foreignaiddiff}). In the appendix, we demonstrate that, strikingly, Secretary Marco Rubio represents a significant outlier compared to other Secretaries of State. In particular, he is especially likely, compared to all other Secretaries, Republican and Democrat, to reference foreign with relative negativity. This patterns coincides with the new administration's decision to abruptly dismantle its longstanding foreign aid arm, USAID \citep{rohner2026aiding}.

\section*{Discussion and Conclusion}

In closing, this paper has introduced the USSOS Speech Dataset, a novel dataset on the public communication of US Secretaries of State between the years 1997 and 2026, covering much of the post-Cold War era. Studying Secretaries of State is important as they serve as the chief foreign affairs advisor to presidents in the United States. While they represent key actors in the context of international relations, in general, and US foreign policy, in particular, much of the literature in political science has overlooked foreign ministers, particularly US Secretaries of State. Instead, existing research opts to focus on the president and executive leaders in other countries as well as members of legislatures, with only a handful of exceptions. We build on the literature of visits by foreign policy actors \citep{malis2021state, mcmanus2025sending, lebovic2016diplomatic, mcmanus2017statements} and shed light on the rhetoric of US Secretaries of State at home \textit{and} abroad. In essence, existing work indicates that key actors, such as bureaucrats, and other actors in the foreign policy space can often diverge in their preferences from those of the president and others in the executive branch, meaning separate actors are not simple proxies for executives \citep{schub2022informing, jost2024advisers}. 

By introducing this dataset, which spans three decades, we are able to track who Secretaries of State talk about, the extent to which these actors talk about different topics, where they deliver remarks, and how this relates to both domestic and international developments. In this paper, we have shown that the USSOS Speech Dataset can be used to answer a range of questions, including the extent to which countries pay attention to foreign aid and how this diverges along partisan lines. More broadly, the dataset can be used to answer a range of critical questions at the intersection of international relations and US foreign policy. 

While scholarship on visits in international relations has grown \citep[e.g.,][]{lebovic2016diplomatic, mcmanus2025sending, malis2021state}, public communication represents a distinct dimension of international diplomacy. Public statements can serve as important signals to other countries, shape expectations about future policy, and potentially constrain leaders' subsequent choices \citep[e.g.,][]{fearon1994signaling, mcmanus2017statements}. In this way, the dataset will enable scholars to assess the determinants and effects of US foreign policy behavior on a range of outcomes. Beyond the dataset's utility for understanding US foreign policy, our work helps spearhead a new research agenda on foreign ministers and their behavior. Future work could, for instance, explore the public discourse of other foreign ministers of leading states and organizations, such as Russia, China, as well as the EU and its members.

\pagebreak

\bibliographystyle{apsr}
\singlespacing{\bibliography{cites.bib}}

\clearpage
\appendix

\setcounter{figure}{0}
\setcounter{table}{0}
\setcounter{page}{1}
\renewcommand{\thefigure}{A\arabic{figure}}
\renewcommand{\thetable}{A\arabic{table}}
\renewcommand{\thepage}{A\arabic{page}}


\section{Dataset Column Descriptions and Quality Checks}
\label{sec:schema}

Each of the rows in the corpus constitutes one public communication by a
sitting U.S.\ Secretary of State. The types of remarks captured in the dataset are as follows: a speech, interview, press availability,
statement, or op-ed. All have currently been released between January~8,~1997 (Madeleine Albright) and
September~1,~2026 (Marco Rubio), spanning nine Secretaries of State and six presidential
administrations. Every row in the dataset carries 19 distinct columns: eight descriptive fields
identifying the event, three parallel text renderings at decreasing levels
of processing, and eight machine-derived enrichment columns. Figure~\ref{fig:pipeline}
summarizes how a row travels from the State Department's web archives to the
released file.

\begin{table*}[h!]
\centering
\small
\caption{Corpus schema by Column Name.}
\label{tab:schema}
\begin{tabular}{@{}llll@{}}
\toprule
\textbf{Column} & \textbf{Type} & \textbf{Produced by} \\
\midrule
\soscol{Row Id}            & integer     & release      \\
\soscol{Secretary}         & categorical (9) & extraction  \\
\soscol{Date}              & ISO date    & finalize    \\
\soscol{Title}             & text        & finalize    \\
\soscol{Speech Type}       & categorical (12) & finalize  \\
\soscol{Location}          & \emph{City, Country} & finalize  \\
\soscol{URL}               & url         & extraction   \\
\soscol{SOS Text}          & text        & finalize      \\
\soscol{Natural Text}      & text        & finalize     \\
\soscol{Raw Text}          & text        & extraction    \\
\soscol{Canonical URL}          & url        & extraction    \\
\midrule
\soscol{SOS Word Count}     & integer    & enrichment    \\
\soscol{Natural Word Count} & integer    & enrichment    \\
\soscol{Row Hash}           & sha256 hex & enrichment    \\
\soscol{Speakers Present}   & list       & enrichment    \\
\soscol{People Mentioned (Natural Text)} & list & enrichment  \\
\soscol{People Mentioned (SOS Text)}     & list & enrichment  \\
\soscol{Countries Mentioned}             & list & enrichment  \\
\soscol{Country ISO Codes Mentioned}     & list & enrichment \\
\bottomrule
\end{tabular}
\end{table*}

All five list-valued columns share the same convention: entries are joined by
``\texttt{;\ }'' (semicolon--space), appear in first-appearance order within
the speech, are deduplicated case-insensitively, and cells contain empty
strings when no content was found.

\subsection{Identity and Provenance Columns}

This subsection includes descriptions on the following: (\soscol{Row Hash}, \soscol{Secretary}, \soscol{Date}, \soscol{Title}, \soscol{Speech Type}, \soscol{Location}, \soscol{URL}, \soscol{Canonical URL}) columns. The constructed intent of said columns is to provide a basis from which their respective rows can be organized for easier provenance-checking and cross-correlation.

\begin{fieldnote}{Row Id}
A sequential integer, $0$ to $n$, renumbered at every release. Deliberately excluded from
\soscol{Row Hash} so that renumbering never masquerades as content changes. Rows are cited
across releases by \soscol{URL} or \soscol{Row Hash}.
\end{fieldnote}

\begin{fieldnote}{Secretary}
The speaking Secretary's last name, drawn from a closed set of nine:
Albright, Powell, Rice, Clinton, Kerry, Tillerson, Pompeo, Blinken, Rubio.
This is the corpus's primary grouping variable, with plans to extend in the future. 
\end{fieldnote}

\begin{fieldnote}{Date}
Release date in ISO \texttt{YYYY-MM-DD}. Dates were cross-checked during
finalization against both the URL's embedded year/month and the dateline
within archived pages.
\end{fieldnote}

\begin{fieldnote}{Title}
The communication's title as published, normalized (site suffixes and
boilerplate stripped). Useful for topic filtering before touching full
text.
\end{fieldnote}

\begin{fieldnote}{Speech Type}
A controlled vocabulary of twelve communication formats: Remarks,
Press Statement, Interview, Statement, Press
Availability, Speech, Press Briefing, Joint Press
Availability, Testimony, Remarks to the Press, Op-Ed,
and Town Hall. We treat speech format to be analytically-separable.
\end{fieldnote}

\begin{fieldnote}{Location}
Where a given communication was delivered. The column follows strict
\emph{City, Country} form (U.S. locations as \emph{City, State}).
Country names are standardized to one spelling corpus-wide. Three non-geographic
values (\texttt{Op-Ed}, \texttt{Recorded}, \texttt{On Travel}) cover
placeless rows.
\end{fieldnote}

\begin{fieldnote}{URL}
The source URL from the State Department's era-partitioned web
archives (e.g., \texttt{1997-2001.state.gov}, \texttt{2021-2025.state.gov}) or active Secretary domain (e.g., \texttt{www.state.gov}).
Beyond provenance, the full URL is the corpus's cross-release citation key. Verified to be back-traceable via direct /\texttt{www}/-link, as determined during initial scraping date of appended rows or transitions to new Secretaries.
\end{fieldnote}

\begin{fieldnote}{Canonical URL}
The URL from the State Department's website (e.g., \texttt{www.state.gov}), as retrieved at the time of scraping.
The Canonical URL is static and serves as a record of before transitions to new Secretaries, as is commonly the case with new presidential administrations.
\end{fieldnote}

\subsection{Textual Content Columns}

This subsection includes descriptions on three deliberately shipped renderings, (\soscol{Raw Text}, \soscol{Natural Text}, \soscol{SOS Text}), for every scraped document. 

\begin{fieldnote}{Raw Text}
The archived page's text exactly as scraped, including: navigation fragments, datelines,
and all. This enables auditing within the pipeline since \soscol{Raw
Text} shows full source content. This column is input from which
\soscol{Natural Text} is derived.
\end{fieldnote}

\begin{fieldnote}{Natural Text}
The full communication with boilerplate removed but \emph{all} voices
retained: speaker labels (\texttt{SECRETARY BLINKEN:},
\texttt{QUESTION:}), moderators, interviewers, and co-speakers, plus stage
directions like \texttt{(Laughter.)}. This is the tier for
conversation-level analysis, enabling measurement on question and answer dynamics, interruption,
co-speaker framing. This column is used as input for the final \soscol{SOS Text} column.
\end{fieldnote}

\begin{fieldnote}{SOS Text}
This is the speech tier for any utterance from just
\emph{the Secretary}, segmented by speaker label.
Non-Secretary turns are dropped, labels are removed, and audience
reactions are stripped. For single-speaker
formats (op-eds, prepared speeches) the whole cleaned document is the
Secretary's. This column was conceptualized given how a press availability with interviewer utterances would contaminate any signal directly relevant to a given Secretary's speech patterns. 
\end{fieldnote}

\subsection{Data Enrichment Columns}

Within the following subsection, we describe the corpus columns, (\soscol{SOS Word Count / Natural Word Count}, \soscol{Row Hash}, \soscol{Speakers Present}, \soscol{People Mentioned (Natural Text) / (SOS Text)}, \soscol{Countries Mentioned / Country ISO Codes Mentioned}), deliberately constructed within the extraction pipeline to enrich information stored within the verified corpus. The columns are in a standardized format to simplify analysis and cross-verifiability for future studies.

\begin{fieldnote}{SOS Word Count / Natural Word Count}
Whitespace-token counts of the two cleaned
tiers. The pair supports instant length normalization and a
cheap structural signal: if both counts are nearly equal the row is
effectively a monologue, while large gaps flag dialogue.
\end{fieldnote}

\begin{fieldnote}{Row Hash}
A SHA-256 digest of the row's nine content fields (all columns except
\soscol{Row Id}), joined with an ASCII unit separator. The hash intentionally
excludes \soscol{Row Id} for stability across new releases. A row's
hash changes if and only if its content changes. This makes silent edits
detectable, allowing replication as a precise content citation.
\end{fieldnote}

\begin{fieldnote}{Speakers Present}
Explicit labels for everyone holding the floor in \soscol{Natural Text}. Organized by parsed speaker label and
ordered by first appearance. Named participants resolve to readable form,
(e.g. \emph{Foreign Minister Lavrov}, \emph{Ambassador Glendon}). Anonymous
roles are kept in brackets, (e.g. \texttt{[QUESTION]},
\texttt{[MODERATOR]}), enabling distinguishment from surnames. The
row's own Secretary is always included, so the column is never
empty. As an example, separate studies can filter
on \texttt{[QUESTION]} for interactive settings, or on foreign-minister
labels for joint appearances. 
\end{fieldnote}

\begin{fieldnote}{People Mentioned (Natural Text) / (SOS Text)}
Any third parties \emph{spoken about}, as extracted by named-entity
recognition over \soscol{Natural Text}~\citep{spacy}. The \emph{(SOS Text)} is a per-row subset of the \emph{(Natural Text)} column.
\end{fieldnote}

\begin{fieldnote}{Countries Mentioned / Country ISO Codes Mentioned}
Place references resolved to canonical country names and ISO 3166-1 alpha-3
codes through a curated gazetteer.
Both columns are strictly positional: the $i$-th code belongs to the
$i$-th name, and entities without an ISO code carry the placeholder
\texttt{N/A}. For example, Kosovo and historical Yugoslavia both carry no
ISO code, whereas Serbia carries \texttt{SRB}.
\end{fieldnote}

\begin{figure}[H]
    \centering
    \includegraphics[width=0.9\textwidth]{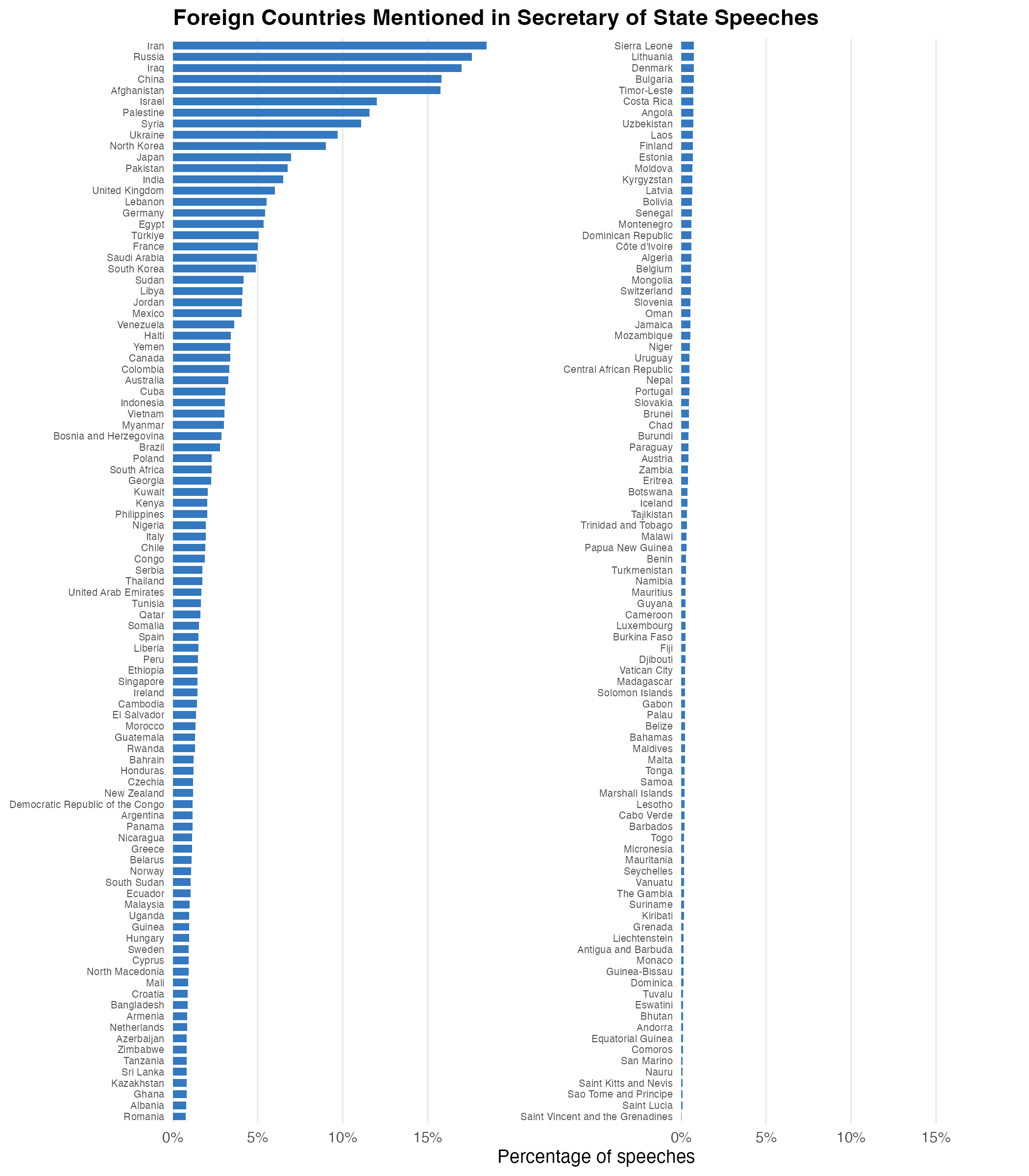}
    \caption{Countries Most Likely to be Mentioned}
    \label{fig:secstatecountries}
\end{figure}

\begin{figure}[H]
    \centering
    \includegraphics[width=\textwidth]{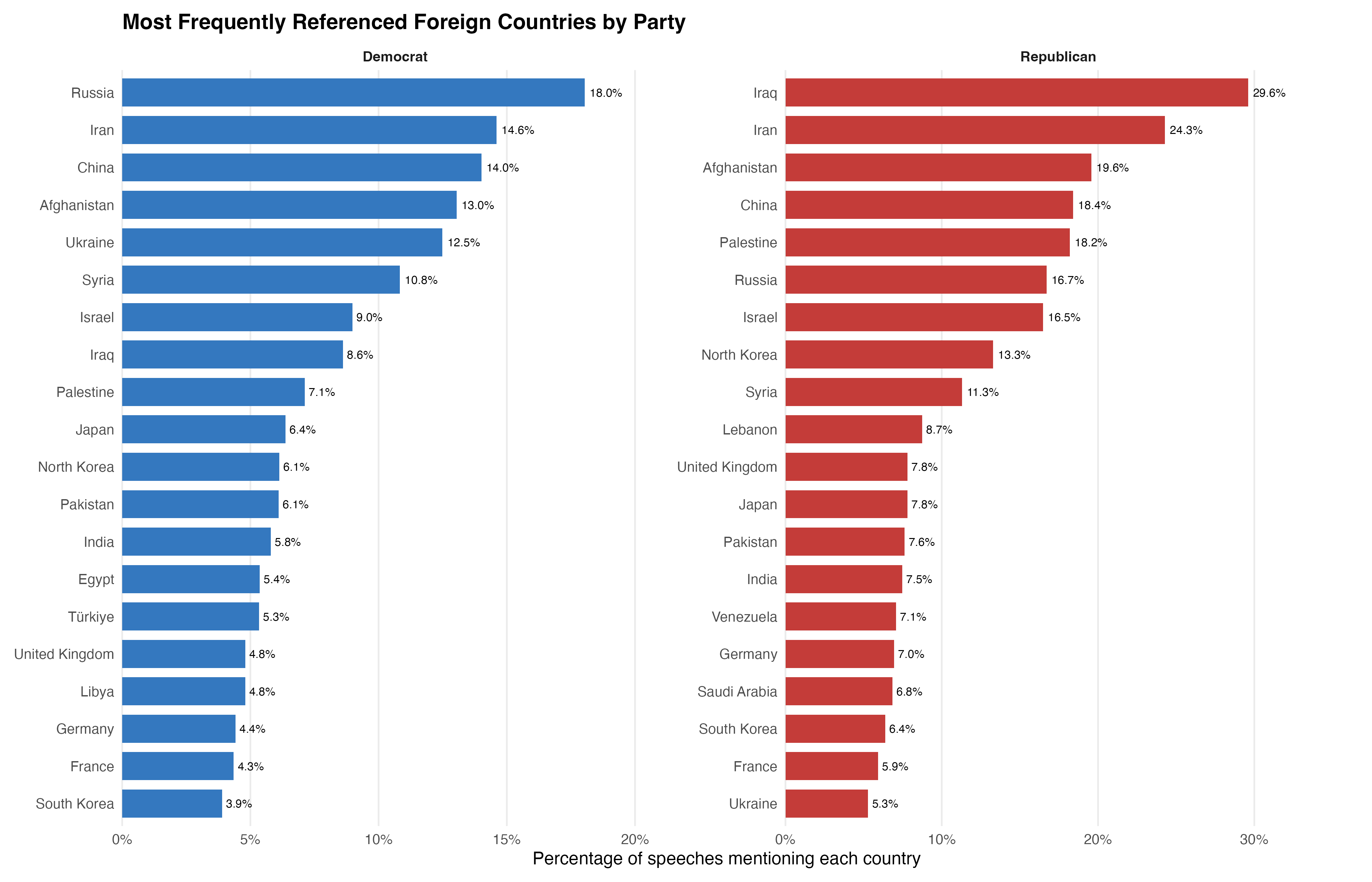}
    \caption{Top 20 Countries Mentioned by Party}
    \label{fig:countriesparty}
\end{figure}

\begin{longtable}{p{0.45\textwidth}p{0.45\textwidth}}
\caption{Countries Never Mentioned by Secretary of State} \\
\toprule
\textbf{Country} & \textbf{Secretaries} \\
\midrule
\endfirsthead
\toprule
\textbf{Country} & \textbf{Secretaries} \\
\midrule
\endhead
\bottomrule
\endfoot
Algeria & Tillerson \\
Andorra & Albright, Powell, Rice \\
Antigua and Barbuda & Albright, Powell, Rice \\
Barbados & Albright, Rice \\
Belize & Rice \\
Bhutan & Albright, Powell, Rice \\
Brunei & Rice \\
Burkina Faso & Albright \\
Cabo Verde & Albright \\
Cameroon & Albright \\
Central African Republic & Rice \\
Comoros & Albright, Clinton, Powell \\
Czechia & Tillerson \\
Djibouti & Albright \\
Dominica & Albright, Powell, Rice \\
Equatorial Guinea & Powell \\
Eswatini & Albright, Powell, Rice \\
Gabon & Albright, Rice \\
Guyana & Rice \\
Kiribati & Albright, Powell, Rice \\
Lesotho & Albright \\
Liechtenstein & Powell, Rice \\
Maldives & Rice, Tillerson \\
Malta & Powell \\
Marshall Islands & Albright, Powell \\
Mauritania & Powell \\
Mauritius & Rice \\
Micronesia & Albright, Powell \\
Monaco & Albright, Powell \\
Morocco & Tillerson \\
Namibia & Powell, Rice, Tillerson \\
Nauru & Albright, Powell, Rice \\
Oman & Tillerson \\
Palau & Albright, Powell \\
Papua New Guinea & Rice \\
Saint Kitts and Nevis & Albright, Pompeo, Powell, Rice \\
Saint Lucia & Clinton, Rice \\
Saint Vincent and the Grenadines & Albright, Clinton, Kerry, Powell, Rice, Tillerson \\
Samoa & Albright, Powell \\
San Marino & Albright, Powell, Rice \\
Sao Tome and Principe & Albright, Powell, Rice \\
Seychelles & Rice \\
Solomon Islands & Albright, Powell \\
South Sudan & Albright, Powell \\
Suriname & Powell, Rice \\
The Gambia & Albright, Powell, Rice \\
Togo & Powell \\
Tonga & Albright, Powell \\
Trinidad and Tobago & Powell, Rice \\
Tuvalu & Albright, Powell, Rice \\
Vatican City & Blinken, Clinton \\
Zambia & Albright
 \label{tab:countriesmentioned}
\end{longtable}

\begin{figure}[H]
    \centering
    \includegraphics[width=0.9\textwidth]{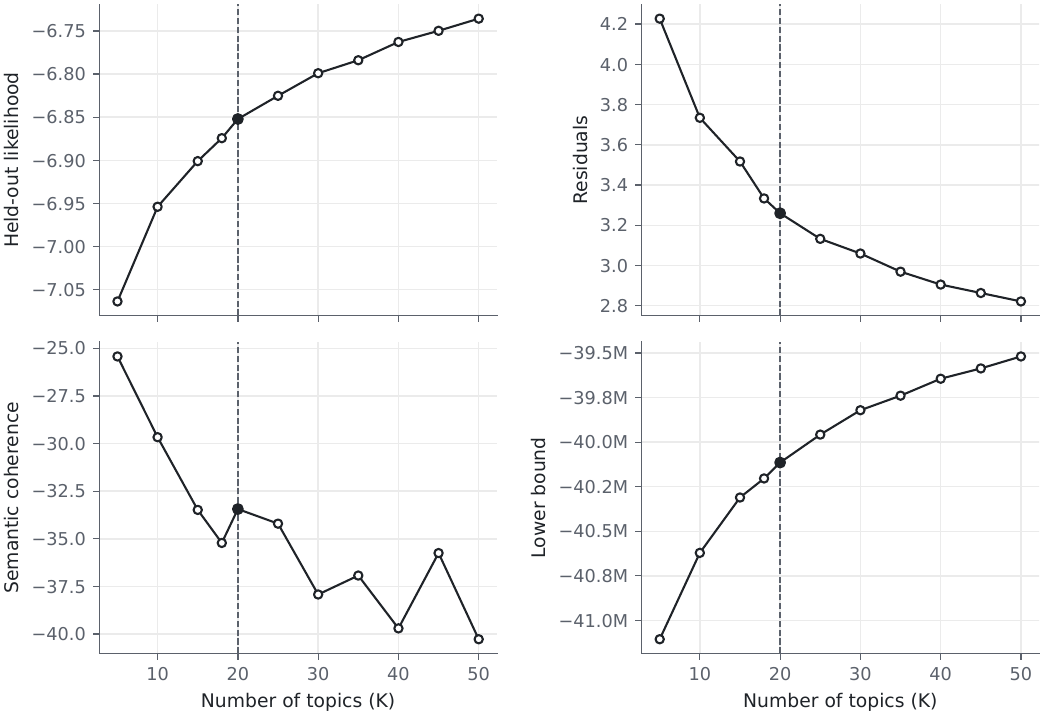}
    \caption{STM Diagnostics}
    \label{fig:diagnostics}
\end{figure}

\begin{figure}[H]
    \centering
    \includegraphics[width=0.85\textwidth]{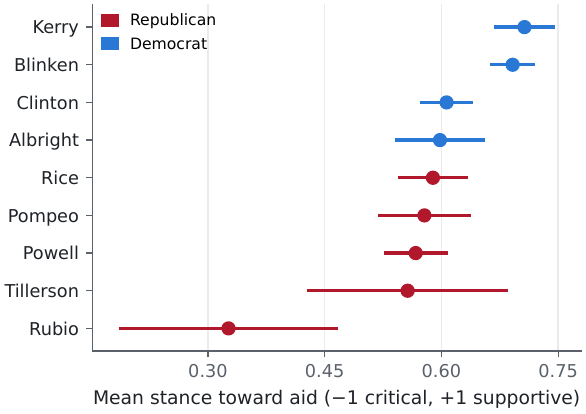}
    \caption{Differences in Foreign Aid Stance By Secretary}
    \label{fig:foreignaiddiff}
\end{figure}

\subsection{Codebook -- System Prompt}
\label{subsec:llm-cb-country}

We applied Qwen3.8-27B as the LLM coder for stance and sentiment. The LLM was loaded on 2 A5000 24GB GPUs with FP8 quantization and thinking set to off. We include the system prompt, split into three sections, for the LLM coding pipeline below. Sections are: 
\begin{enumerate}
    \item \textbf{Preamble}, to provide initial guidance to the LLM, 
    \item \textbf{Stance Determination} to indicate a the integer value assigned on whether the provided Secretary of State utterances are: positive = +1, neutral/descriptive = 0, or negative = -1, 
    \item \textbf{Rules}, 10 labeled examples which indicate the explicit threshold for whether the LLM decides if a provided utterance is positive, neutral, or negative
\end{enumerate}

\begin{soscodebook}{Country Stance Codebook}{System Prompt: Preamble}
\label{box:cb-country}

{\color{black}
You will read one \textbf{target sentence} spoken by a U.S. Secretary of State and the name of one \textbf{target country} mentioned in it. The sentence before and after it are shown for context only. Code one thing: the stance the Secretary takes in the target sentence toward the \textbf{government of the target country}, that is, the state as an international actor (its leaders, regime, policies, and conduct).
\par\smallskip
}

\end{soscodebook}

\begin{sosscale}{Stance Determination}
\begin{tabularx}{\linewidth}{@{}c>{\raggedright\arraybackslash}p{2.3cm}>{\raggedright\arraybackslash}X@{}}
\toprule
\textsf{\scriptsize Code} & \textsf{\scriptsize Label} & \textsf{\scriptsize Definition and examples} \\
\midrule

\soscode{soscat3}{+1} & \textbf{Positive / Favorable} & treats the government as a partner, ally, or friend. This includes praising, thanking, or congratulating it; endorsing its policies or conduct; presenting cooperation with it as valued; and expressing support for, or solidarity with, the government (\textit{``Japan is a cornerstone of peace and stability in the Pacific.''} \textit{``We thank the Government of Jordan for its leadership.''} \textit{``We stand with Ukraine as it defends its sovereignty.''}). \textbf{Ceremonial goodwill is favorable} (see the first rule below). \\ \addlinespace

\soscode{sosgray}{0} & \textbf{Neutral / Descriptive} & mentions the country without judging its government. This covers travel, meetings, agendas, places and venues, lists of participants, factual reports, and the country as a mere location (\textit{``I will travel to Madrid next week.''} \textit{``I met with the Foreign Minister of Brazil.''} \textit{``The talks were held in Geneva.''}). \\ \addlinespace

\soscode{soscat2}{$-$1} & \textbf{Negative / Not Favorable} & treats the government as an adversary, problem, or threat. This includes condemning, criticizing, or accusing it; demanding that it change course; announcing or defending sanctions or pressure against it; and calling it a \textit{``regime''}, aggressive, repressive, corrupt, or untrustworthy (\textit{``Iran continues to be the world's leading state sponsor of terrorism.''} \textit{``We call on Beijing to release those unjustly detained.''} \textit{``Russia must withdraw its forces.''}). \\

\bottomrule
\end{tabularx}
\end{sosscale}

\begin{sosrules}{Rules}
\begin{enumerate}\setlength{\itemsep}{2pt}
\item \textbf{Ceremonial goodwill toward the country is +1, even when formulaic.} This covers congratulations or best wishes to the country, its people or its leaders (national day, independence day, anniversary, holiday, election, new government); condolences and sympathy after a disaster, attack or death; and thanks or welcome extended to the country, its people or its leaders. These messages are a deliberate diplomatic signal of goodwill toward the state (\textit{``On behalf of the American people, I congratulate the people of Ghana on their national day.''} $\rightarrow$ Ghana +1; \textit{``Our thoughts are with the people of Turkey after this terrible earthquake.''} $\rightarrow$ Turkey +1). The exception is the next rule.
\item Code toward the \textbf{government/state}, not the people. Support for a country's people \emph{against} its government is \textbf{$-$1} (\textit{``We stand with the Iranian people in their call for freedom.''} $\rightarrow$ Iran $-$1). Goodwill addressed to a people \emph{in opposition to} their government is not ceremonial goodwill: \textit{``We congratulate the people of Venezuela on their courage in standing up to the Maduro regime''} $\rightarrow$ Venezuela $-$1.
\item Code \textbf{only the target country}. Criticism of a third party is not unfavorable to the target. In \textit{``We condemn Russia's invasion of Ukraine''}, Russia is $-$1. Ukraine is 0, or +1 if support for it is expressed.
\item A \textbf{negative situation} inside the country (war, famine, crisis) is not by itself unfavorable to its government. Code $-$1 only when the government is blamed or pressured.
\item \textbf{Mixed} sentences (\textit{``We have differences with China, but we cooperate on climate''}): code the stance the sentence ends on or stresses. If the two parts are truly balanced, code 0.
\item \textbf{Engagement is not endorsement.} \textit{``We are talking to the Russians about arms control''} is 0. It is +1 only if the relationship or cooperation is praised or valued.
\item A \textbf{demand phrased politely} is still a demand: \textit{``We urge the Government of Egypt to release the journalists''} is $-$1.
\item A \textbf{joint statement or list of countries} with no judgment of any single one is 0 for each.
\item Use context only to resolve what \textit{``they''}, \textit{``this government''}, and similar references point to. Code the target sentence itself.
\item If you genuinely cannot tell, choose 0 and tick \textbf{unsure}.
\end{enumerate}
\end{sosrules}

\subsection{Correlation Analysis to UN Voting Alignment}

\begin{figure}[H]
    \centering
    \includegraphics[width=0.99\textwidth]{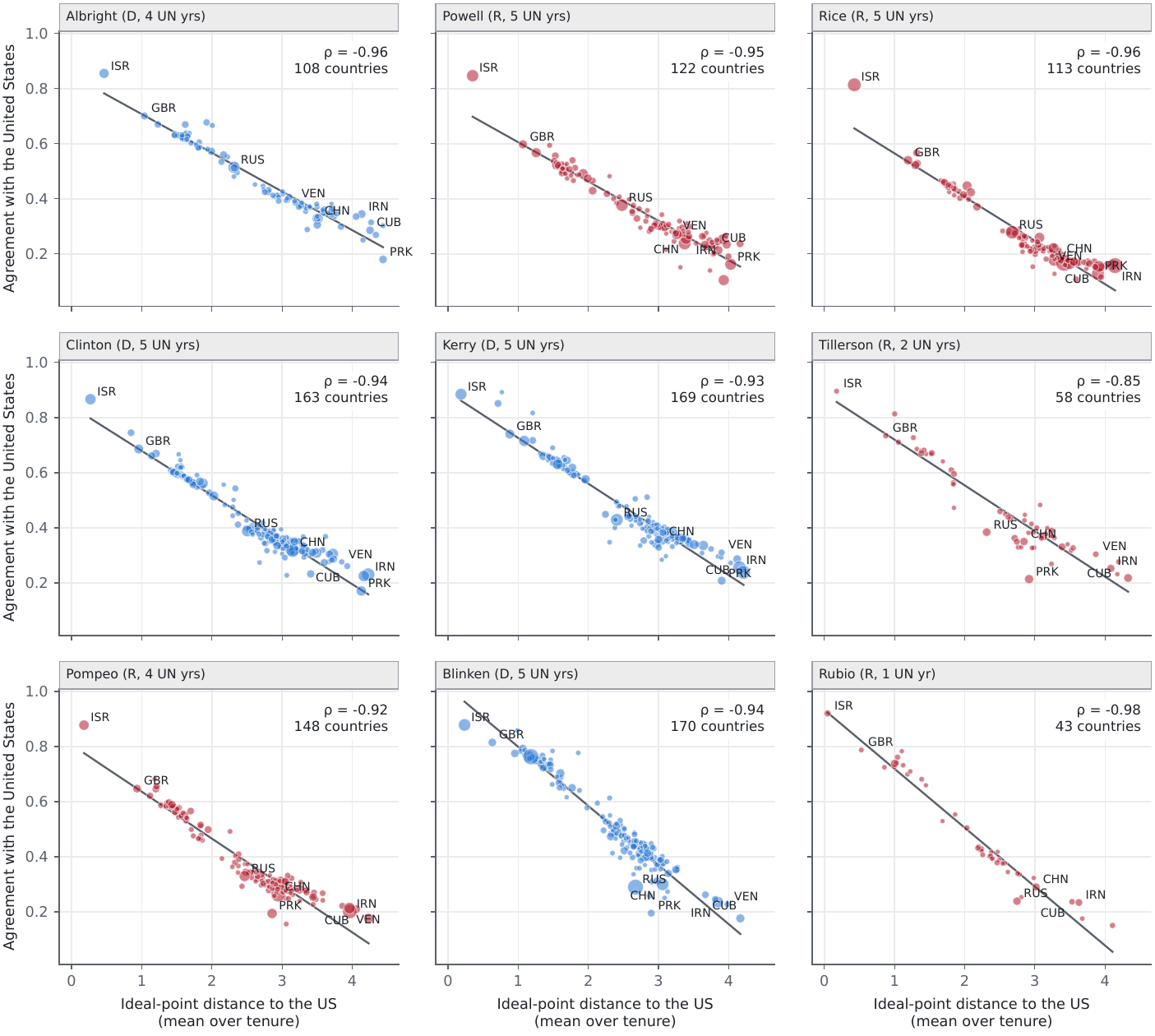}
    \caption{Differences in Foreign Aid Stance By Secretary}
    \label{fig:unga_by_sec}
\end{figure}

The following subsection continues the investigation in the correlation between the UN General Assembly voting alignment measure by \citep{bailey2017estimating} and a per-secretary breakdown of their given utterances alignment with the United States on a per-country basis. Each scatter point within Figure~\ref{fig:unga_by_sec} represents a unique country described from the Secretary of State during their tenure. Therefore, the line indicates the general trend of the Secretary of State's rhetoric (following Figure~\ref{fig:ungavoting} in the main paper). Additional details within Figure~\ref{fig:unga_by_sec} show: mean correlation ($\rho$) of the trend, the number of unique countries mentioned during the Secretary of State's tenure, and highlighted countries along decreasing points in the trend line. In short, the correlation is stable across Secretaries.

\subsection{Median Spearman Correlation of SoS Rhetoric Over Time}

\begin{figure}[H]
    \centering
    \includegraphics[width=0.99\textwidth]{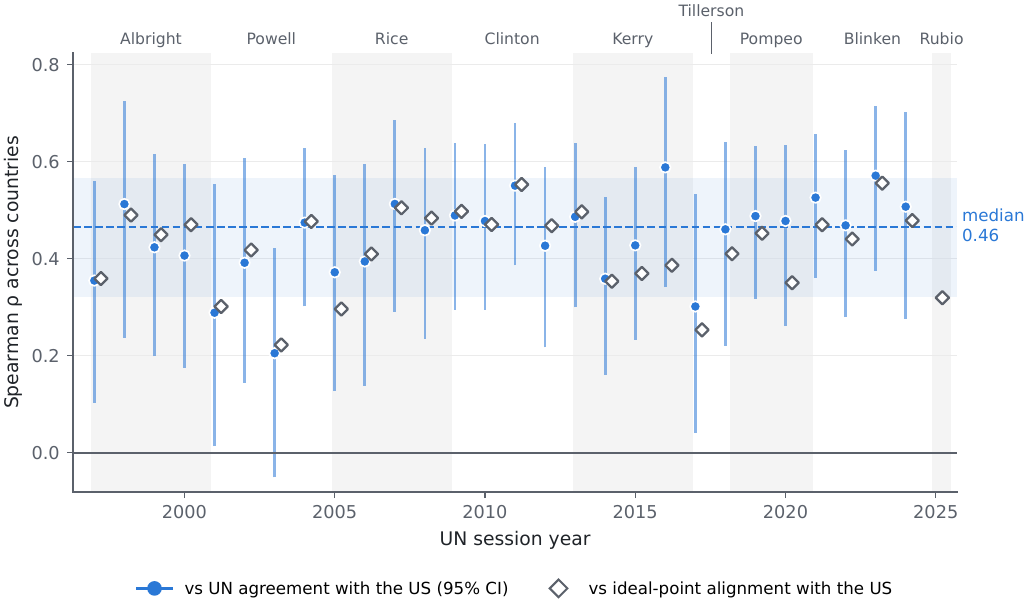}
    \caption{Median Spearman Correlation of SoS Rhetoric Over Time}
    \label{fig:temporal_sos}
\end{figure}

The following subsection is an extension of the analysis demonstrated with Figure~\ref{fig:ungavoting}. Here, Figure~\ref{fig:temporal_sos} indicates the median spearman correlation for rhetoric from each Secretary of State during their tenure, relative to how close the United States delegation vote aligns with the other UN delegate country votes (shown in blue) and how the Secretary of State's rhetoric aligns with the Voeten~\cite{bailey2017estimating} ideal-point ranking between the US and UN delegate countries (shown in white). 

\subsection{How the Corpus Is Verified}

Given that this corpus is updated automatically, we construct an aggressive suite of verification checks. These checks operate in three stages. The first, operates after text-tier extraction as 'Pre-Quality Checks'. The initial pre-quality checks assert that appending and enriching new speeches to the pre-existing corpus match quality standards. We ensure these standards via quality checks sorted into 26 distinct categories, labeled in alphabetical order (A-to-Z). A non-exhaustive list of issues addressed by the checks are: 
\begin{itemize}
    \item Boilerplate and navigation residue appearing like speech
    \item Incorrect speaker attribution
    \item Duplicated rows under different URLs
    \item Date contradiction within source pages
    \item Identified scrape failures
    \item Failed agreement between or within enrichment columns
\end{itemize}

The (A-to-Z) checks are repeated after enrichment as a final 'Verification' step. The verification step adds schema and identity preamble as well as secretary coverage scans, verifying no contamination from the weekly merge and enriched data. The schema and identity preamble identifies potential discrepancies between already-mapped row values and the incoming update rows. Finally, the per-secretary coverage scan is an $N$-gram analysis, ($N$ = \{$2-10$\}), over incoming row content. The $N$-gram analysis surfaces anomalies in updated rows, often catching changes in archival techniques between administrations. We document each check individually within Table~\ref{tab:checks}.


\subsection{Verification Check Details}
\label{sec:verification}

Before publishing, every corpus release must pass $+83$ verification checks, labeled with 'Blocking' severity. Informative checks are labeled as 'Advisory', since their results require human inspection to determine whether action is required. The pipeline runs automatically every week, executing checks
over \emph{every incoming} row. Table~\ref{tab:checks}
provides details on every check within the automated corpus pipeline (up-to-date as of September 1, 2026). Each verification check was constructed to address issues in observable categories from months of dataset construction that involved the time and labor of the authors.

\begin{table*}[!h]
\centering
\scriptsize
\caption{Verification Check Table}
\label{tab:checks}
\begin{tabular}{@{}cll@{}}
\toprule
\textbf{Check} & \textbf{Scenario Requiring Guard} & \textbf{Severity} \\
\midrule
\multicolumn{3}{@{}l}{\emph{Data Access}} \\
O1 & Response from Secretary of State URL & Blocking \\
\multicolumn{3}{@{}l}{\emph{Schema and Row Identity}} \\
P1 & Missing columns, gaps in row numbering, blank fields & Blocking \\
P2 & Misspelled secretary name or speech format splitting a category & Blocking \\
P3 & Same speech stored twice, or a date its own source page contradicts & Blocking \\
P4 & Website furniture and speaker labels left inside speech text & Blocking \\
\addlinespace
\multicolumn{3}{@{}l}{\emph{Text Cleanliness}} \\
A & Page headers stored as though it were a speech & Blocking \\
B & Site footer left in the text & Blocking \\
C & Old navigation bar left in the text & Blocking \\
D & A press-release date stamp left in the body & Blocking \\
E & Document filing codes left at the start or end of the text & Blocking \\
F & Audience reactions counted as the Secretary's words & Blocking \\
G & Transcriptionist notes counted as the Secretary's words & Blocking \\
H & Incoherent character patterns from transcription typos & Blocking \\
I & Encoding errors which cause 'mojibake' characters & Blocking \\
J & Unusually short speeches that may need a length floor & Advisory \\
AA & .gov business sidebar and translation-link header cleanup  & Blocking \\
AB & "ã——" paragraph-separator encoding error removal & Blocking \\
\addlinespace
\multicolumn{3}{@{}l}{\emph{Descriptive Fields}} \\
K & Op-Ed placeholder appearing where a real location belongs & Blocking \\
L & A location value missing its country & Blocking \\
M & A speech by a deputy filed under the Secretary & Blocking \\
N & Both Congos rendered so they cannot be told apart & Blocking \\
O & ``En Route'' standing in for a real place & Blocking \\
P & A venue name padding out the location field & Blocking \\
\addlinespace
\multicolumn{3}{@{}l}{\emph{Enrichment Integrity}} \\
Q & The enrichment step silently rewriting the speech text & Blocking \\
R & Word counts that disagree with the text they describe & Blocking \\
S & Altered content, duplicate content, or a stale manifest & Blocking \\
T & A speaker list missing its own Secretary or confusing roles with names & Blocking \\
U & Speaker lists with implausibly short, long, or numerous entries & Advisory \\
\addlinespace
\multicolumn{3}{@{}l}{\emph{Machine-Extracted Columns}} \\
V & People lists with duplicates, stray labels, or impossible membership & Blocking \\
W & country names misaligned with their ISO codes & Blocking \\
X & Encoding damage or non-names in the extracted columns & Blocking \\
Y & Same person recorded under two spellings & Advisory \\
\addlinespace
\multicolumn{3}{@{}l}{\emph{Coverage Testing}} \\
Z & A failed scrape leaving a silent hole in a secretary's record & Blocking \\
$N$-grams & Newly-filtered rows contain boilerplate, anomalous or unmapped values & Advisory \\
\bottomrule
\end{tabular}
\end{table*}

\pagebreak


\section{Worked Examples from Corpus Rows}
\label{sec:usecases}

Below, we include Figures~\ref{fig:across} and \ref{fig:network} as a source of inspiration for future study into Secretary of State speech dynamics, both of which can emerge from using the dataset. Each figure below is built from a specific row contained within the corpus. 

\begin{figure*}[h]
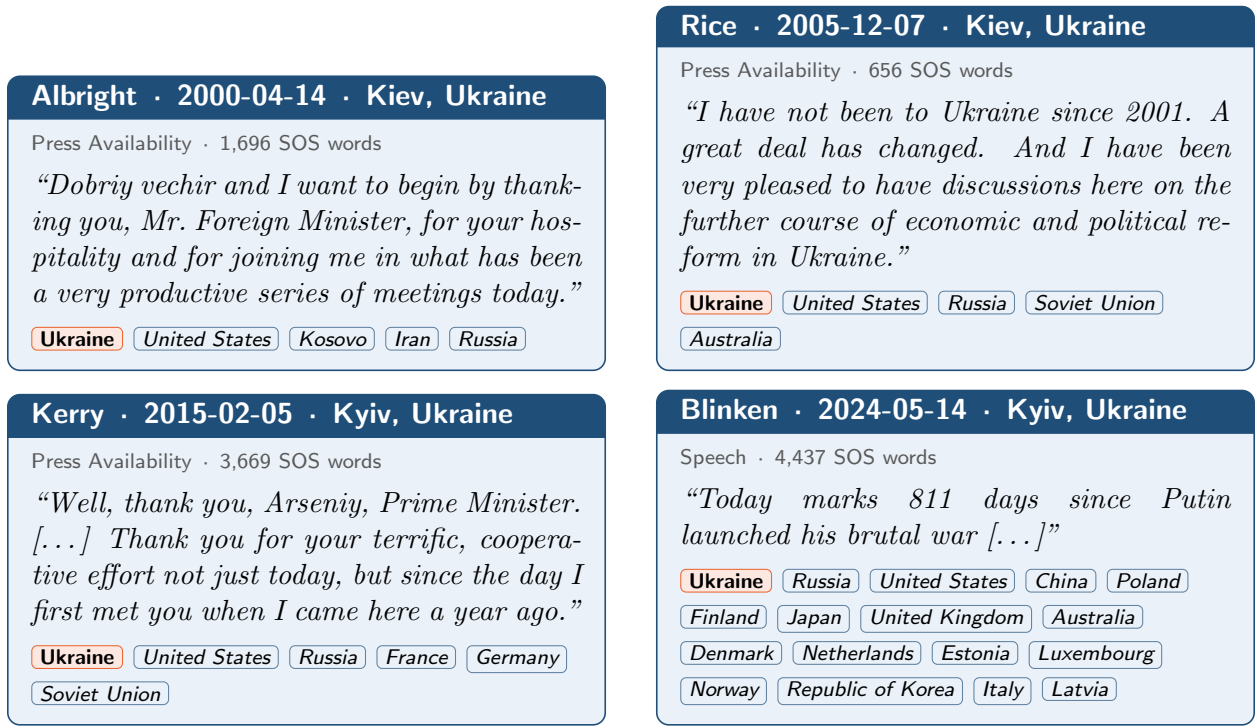

\centering
\begin{minipage}[t]{0.48\textwidth}
\begin{rowcard}{Albright \,\textperiodcentered\, 2000-04-14 \,\textperiodcentered\, Kiev, Ukraine}
\scriptsize \textcolor{sosgray}{\textsf{Press Availability \,\textperiodcentered\, 1{,}696 SOS words}}\par\smallskip
\small\itshape ``Dobriy vechir and I want to begin by thanking you, Mr.\
Foreign Minister, for your hospitality and for joining me in what has been a
very productive series of meetings today.''\par\smallskip
\raggedright
\chiphl{Ukraine} \chip{United States} \chip{Kosovo} \chip{Iran} \chip{Russia}
\end{rowcard}
\end{minipage}\hfill
\begin{minipage}[t]{0.48\textwidth}
\begin{rowcard}{Rice \,\textperiodcentered\, 2005-12-07 \,\textperiodcentered\, Kiev, Ukraine}
\scriptsize \textcolor{sosgray}{\textsf{Press Availability \,\textperiodcentered\, 656 SOS words}}\par\smallskip
\small\itshape ``I have not been to Ukraine since 2001. A great deal has
changed. And I have been very pleased to have discussions here on the further
course of economic and political reform in Ukraine.''\par\smallskip
\raggedright
\chiphl{Ukraine} \chip{United States} \chip{Russia} \chip{Soviet Union} \chip{Australia}
\end{rowcard}
\end{minipage}

\medskip

\begin{minipage}[t]{0.48\textwidth}
\begin{rowcard}{Kerry \,\textperiodcentered\, 2015-02-05 \,\textperiodcentered\, Kyiv, Ukraine}
\scriptsize \textcolor{sosgray}{\textsf{Press Availability \,\textperiodcentered\, 3{,}669 SOS words}}\par\smallskip
\small\itshape ``Well, thank you, Arseniy, Prime Minister. [\dots] Thank you
for your terrific, cooperative effort not just today, but since the day I
first met you when I came here a year ago.''\par\smallskip
\raggedright
\chiphl{Ukraine} \chip{United States} \chip{Russia} \chip{France} \chip{Germany} \chip{Soviet Union}
\end{rowcard}
\end{minipage}\hfill
\begin{minipage}[t]{0.48\textwidth}
\begin{rowcard}{Blinken \,\textperiodcentered\, 2024-05-14 \,\textperiodcentered\, Kyiv, Ukraine}
\scriptsize \textcolor{sosgray}{\textsf{Speech \,\textperiodcentered\, 4{,}437 SOS words}}\par\smallskip
\small\itshape ``Today marks 811 days since Putin launched his brutal war
[\dots]''\par\smallskip
\raggedright
\chiphl{Ukraine} \chip{Russia} \chip{United States} \chip{China} \chip{Poland}
\chip{Finland} \chip{Japan} \chip{United Kingdom} \chip{Australia} \chip{Denmark} \chip{Netherlands}  \chip{Estonia} \chip{Luxembourg} \chip{Norway} \chip{Republic of Korea} \chip{Italy} \chip{Latvia}
\end{rowcard}
\end{minipage}
\caption{\textbf{References to Ukraine Across Secretaries.}
Four dated rows, where a Secretary of State speaks in Ukraine's capital
alongside a Ukrainian counterpart. 
``Kyiv'.}
\label{fig:across}
\end{figure*}

\begin{figure*}[h]
\centering
\begin{tikzpicture}[
  font=\scriptsize\sffamily,
  sec/.style={draw=sosblue, fill=sosblue, text=white, rounded corners=2pt,
              inner sep=3pt, font=\small\sffamily\bfseries},
  spk/.style={draw=sosblue, fill=sosbluepale, rounded corners=2pt, inner sep=2.5pt},
  anon/.style={draw=sosgray, fill=white, rounded corners=2pt, inner sep=2.5pt,
               font=\scriptsize\ttfamily, text=sosgray},
  ppl/.style={draw=soscat2, fill=soscat2!12, rounded corners=2pt, inner sep=2.5pt},
  edge/.style={sosblue!50, thick},
  pedge/.style={soscat2!70, thick, dashed}]

\node[sec] (rice) at (0,0) {Rice};

\node[spk] (straw)  at (-4.4, 1.5) {Foreign Secretary Straw};
\node[spk] (db)     at (-4.2,-0.8) {FM Douste-Blazy};
\node[spk] (mach)   at (-1.6,-1.9) {FM Machumura};
\node[spk] (lav)    at ( 1.6,-1.9) {FM Lavrov};
\node[spk] (ass)    at ( 4.2,-0.8) {FM Asselborn};
\node[anon] (mod)   at ( 4.6, 0.6) {[MODERATOR]};
\node[anon] (qq)    at ( 4.0, 1.7) {[QUESTION]};

\foreach \s in {straw, db, mach, lav, ass, mod, qq}
  \draw[edge] (rice) -- (\s);

\node[ppl] (abbas) at (0, 2.1) {Mahmoud Abbas};
\draw[pedge] (rice) -- (abbas);

\node[draw=sosblue!50, fill=white, rounded corners=2pt, inner sep=4pt,
      text width=112mm, align=flush left, anchor=north, font=\scriptsize]
  (countries) at (0,-2.9)
  {\textcolor{sosblue}{\textsf{\bfseries Countries Mentioned (23):}}\ \
   United Kingdom \textperiodcentered\ Afghanistan \textperiodcentered\ Israel
   \textperiodcentered\ Gaza \textperiodcentered\ Iran \textperiodcentered\
   France \textperiodcentered\ Germany \textperiodcentered\ Sudan
   \textperiodcentered\ North Korea \textperiodcentered\ Iraq
   \textperiodcentered\ Lebanon \textperiodcentered\ Syria \textperiodcentered\
   Haiti \textperiodcentered\ Zimbabwe \textperiodcentered\ United States
   \textperiodcentered\ Belgium \textperiodcentered\ Turkey \textperiodcentered\
   Saudi Arabia \textperiodcentered\ Jordan \textperiodcentered\ Egypt
   \textperiodcentered\ West Bank \textperiodcentered\ Palestine
   \textperiodcentered\ Luxembourg};
\end{tikzpicture}
\caption{\textbf{Transcript Interaction Network:} Rice's June 2005 joint press availability with G-8 foreign ministers in London.
\textit{Solid Edges:} co-present speakers from \soscol{Speakers Present},
including bracketed anonymous-roles (i.e. [QUESTION]). \textit{Dashed Edge:} A third party, Mahmoud Abbas, named by Secretary Rice within \soscol{People Mentioned (SOS Text)}, independent of the 14 people present in transcript. \textit{Below:} 23 countries resolved from this row. Interaction networks compiled from the full dataset could span co-presence and co-mentions tied between every Secretary.}
\label{fig:network}
\end{figure*}
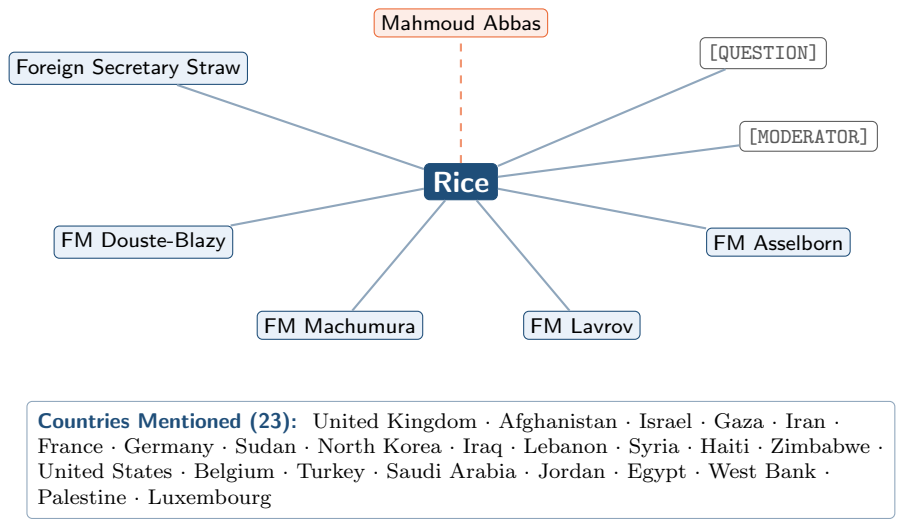

\end{document}